# Crystal Shape and Lattice Deformation in Powder Diffraction

Authors

**Matteo Leoni[a]* and Alberto Leonardi[bcd]***

[a]Research and Analytical Services Department, Saudi Aramco, PO Box 62, Dhahran, 31311, Saudi Arabia

[b]Physical Sciences, Diamond Light Source (United Kingdom), Diamond House - Harwell Science & Innovation Campus, Didcot, Oxfordshire, OX11 0DE, United Kingdom

[c]Earth and Atmospheric Sciences Department, Indiana University Bloomington, 1001 E 10th St, Bloomington, Indiana, IN 47408, USA

[d]Technische Fakultät, Friedrich-Alexander-Universität Erlangen-Nürnberg, Cauerstrasse 3, Erlangen, Bavaria, 91058, Germany

Correspondence email: matteo.leoni@gmail.com; alberto.leonardi@diamond.ac.uk

**Synopsis** This work enables powder diffraction analysis of crystal shape and lattice deformation beyond predefined shape models by treating both effects as coupled yet independently refinable degrees of freedom. This capability is achieved by extending the established common volume function formalism, allowing unified refinement of shape, lattice deformation, and their relative misorientation.

**Abstract** Accurate modelling of diffraction peak shapes is essential for extracting microstructural information from nanocrystalline materials. Common-volume functions (CVFs) are widely used to describe finite-size and shape broadening in powder diffraction, but analytical expressions are available only for a limited set of ideal geometries. Here, we introduce a generalized Fourier-based framework in which crystal-domain shape deformation, lattice deformation, and relative shape–lattice misorientation are treated as independently refinable tensor operations within a unified formalism. The approach enables continuous affine transformations of both crystal shape and lattice base while preserving analytical evaluation of directional Fourier coefficients. As a result, complex particle shapes, anisotropic deformations, and arbitrary relative orientations between shape and lattice can be modelled within a single reciprocal- and real-space framework, including coupled shape–lattice transformations not accessible using conventional powder diffraction line-profile methods. The formalism can be applied to individual diffraction peaks, full powder patterns, and total-scattering shape corrections. Validation against virtual scattering experiment data demonstrates that crystal size, shape, lattice deformation, and relative shape–lattice orientation can be simultaneously recovered with high accuracy.

## 1. Introduction

Nanocrystals underpin a wide range of modern and emerging technologies, with their functional behaviour governed not only by atomic-scale structure but also by crystal shape and its coupling to structure. Over the past decades of fundamental innovation, shape-controlled nanocrystal synthesis is approaching industrial application (Li *et al.*, 2023; Nguyen *et al.*, 2023; National Science Foundation, 2024). In many systems, however, functional properties arise from a coupled interplay between structure and shape deformation, rather than from either factor alone (Boukouvala *et al.*, 2021; Crisp & Singh, 2024). Experimental studies have also shown that nanocrystal morphologies can evolve substantially under chemical and physical stimuli, including thermal annealing, surface diffusion, strain relaxation, and solid–solid transformations, directly linking morphological evolution to underlying structural rearrangements (Fichthorn & Yan, 2021; Waitz *et al.*, 2009; Du *et al.*, 2017; Wang *et al.*, 2016).

Characterization of large nanoparticle ensembles requires techniques capable of probing both structure and shape with statistical relevance. Transmission electron microscopy (TEM), particularly high-resolution TEM, provides direct real-space insight into individual nanocrystals and is indispensable for identifying crystal structure, defects, and local surface terminations. Recent advances in automated and machine-learning-assisted image analysis have further improved throughput and reproducibility, enabling statistically meaningful population-level analysis of size and shape distributions (Aviles & Lear, 2025; Neumann & Rafaja, 2024; Cho *et al.*, 2024). However, TEM-based approaches remain fundamentally limited in capturing coupled shape–lattice evolution across ensembles. While imaging provides detailed local structural and morphological information, it does not directly reveal how shape anisotropy and lattice deformation co-evolve or correlate across reciprocal space. In particular, subtle shape-induced lattice rotations or coupled deformation modes are difficult to quantify reliably, especially in size-dispersed ensembles or systems with concurrent strain relaxation and surface reconstruction.

Despite these advances in synthesis and characterization, systematically distinguishing shape effects from lattice deformation remains a central challenge in nanoscience. Powder diffraction, by contrast, is intrinsically sensitive to such coupled effects through their influence on peak broadening, asymmetry, and profile shape. Nevertheless, conventional diffraction models generally require predefined particle shapes and fixed shape–lattice orientation relationships, preventing direct recovery of coupled deformation and relative shape-lattice rotation from ensemble-averaged data. This problem is particularly acute in quantitative diffraction analysis, where crystal shape and lattice distortions are inherently coupled. Most analytical approaches still treat them as independent contributions, implicitly assuming a fixed relative orientation between crystal shape and crystallographic axes. This assumption limits the ability to quantify shape rotations, facet distortions, and anisotropic reshaping relative to the underlying lattice. As a result, shape–lattice misorientation and coupled deformation modes remain largely inaccessible within standard diffraction-based frameworks, highlighting the need for methodologies that treat shape and lattice structure as independent, yet interacting, degrees of freedom.

The modelling of anisotropic peak broadening in powder diffraction has been approached through several complementary frameworks. Conventional reciprocal-space strain models, such as the Stephens formalism (Stephens, 1999), describe anisotropic broadening phenomenologically via strain tensors within Rietveld refinement, but do not explicitly represent finite crystallite shape or its orientation relative to the lattice. Classical size–strain methods, including the Scherrer equation and Williamson–Hall analysis (Scherrer, 1918; Scardi et al., 2004; Williamson & Hall, 1953; Ungár et al., 1998; Brandstetter et al., 2008; Khorsand Zak & Hashim, 2025; Mohan et al., 2024), provide simple estimates of average crystallite size but inherently conflate size and strain contributions. Spherical-harmonics-based approaches, such as the Popa formalism (Popa, 1998), offer more flexible anisotropic descriptions, yet their coefficients lack direct geometric interpretation in terms of explicit shape or lattice deformation. Overall, these methods rely on strong simplifying assumptions and often conflate size and lattice-distortion effects, while more physically grounded treatments have been developed (Scardi *et al.*, 2004; Ungár *et al.*, 1998; Brandstetter *et al.*, 2008).

More physically grounded approaches include real-space and Fourier-based methods. Pair distribution function (PDF) analysis (Egami & Billinge, 2003) probes local atomic structure but has comparatively limited direct sensitivity to morphology and finite-size shape effects. In contrast, analytical approaches based on the common volume function (CVF) and Fourier-based methods such as Warren–Averbach analysis (Warren & Averbach, 1950, 1952) and Whole Powder Pattern Modelling (WPPM) (Scardi & Leoni, 2002; Leoni, 2019*b*; Scardi, 2020) explicitly connect finite crystallite shape to diffraction profiles. However, these methods are typically restricted to idealized geometries or fixed shape–lattice orientation relationships. Extended approaches combining diffraction and total scattering, including whole PDF modelling (WPDFM) (Leonardi, 2021), extend applicability across length scales but still rely on constrained structural assumptions.

At the atomistic level, the Debye scattering equation (DSE) (Debye, 1915) removes explicit assumptions on shape and lattice symmetry, but its computational cost scales quadratically with atom number, limiting its practical use in iterative refinement of diffraction profiles. Fourier- and CVF-based approaches such as WPPM and WPDFM reduce computational complexity by exploiting reciprocal–real space relationships, but still inherit constraints related to predefined crystal shape and fixed lattice reference frames.

Across all of these approaches, a common limitation persists: crystal shape, lattice deformation, and their relative orientation are not treated as independently variable quantities. Most models implicitly impose a fixed coupling between shape and crystallography or rely on predefined geometries selected a priori. As a result, cases in which shape deformation and lattice deformation evolve simultaneously and with independent orientation cannot be uniquely resolved from powder diffraction data.

To overcome these limitations, Fourier-based approaches have been extended to describe a range of crystallite shapes, including simple shapes such as spheres, cubes, cylinders, tetrahedra, and octahedra, which are frequently used as reference models in diffraction analysis (Scardi & Leoni, 2001; Leoni, 2019*a*). Beyond these closed-form solutions, numerical approaches have been developed for arbitrary polyhedral and complex shapes, including hollow or irregular particles (Leonardi *et al.*, 2012). Building on these ideas, Allegra and Wilson introduced a framework in which diffraction profiles are derived from linear transformations applied to reference shapes, enabling analytical construction of ellipsoidal and parallelepiped domains from spherical and cubic references . Although widely revisited and extended (Balic-Zunic & Dohrup, 1999; Ectors *et al.*, 2015), these approaches still rely on a fixed relationship between crystal lattice and shape axes, and typically treat shape as a rigid or preselected entity rather than a continuously deformable variable.

This limitation is particularly restrictive for modern nanomaterials, where morphology can evolve continuously under growth, strain, or phase transformation conditions. Here, we introduce a unified framework that provides an intermediate description between phenomenological anisotropic broadening models and fully atomistic scattering calculations. The central idea is the introduction of distinct affine operators acting separately on (i) the crystallographic lattice and (ii) the finite crystallite domain within a common Fourier/CVF formalism. This formulation enables crystal shape deformation, lattice deformation, and their relative misorientation to be treated as separately refinable yet coupled parameters while preserving the analytical tractability of directional Fourier coefficients.

By extending the linear transformation concepts originally proposed by Allegra and Wilson, we examine how powder diffraction profiles evolve under independent deformations of (i) domain shape, (ii) lattice geometry, or (iii) both simultaneously. As an example, starting from cubic reference domains and lattices, the framework naturally extends to parallelepiped domains in cubic lattices, cubic domains in triclinic lattices, and fully coupled deformations involving both shape and lattice transformations. This separation of shape and crystallographic symmetry, together with their relative orientation in space, allows diffraction profiles to be expressed in a locally defined reciprocal-space representation, facilitating integration into WPPM and WPDFM-based refinement schemes. Hence, the proposed formalism can be readily implemented in existing whole-pattern analysis software for PD line profile analysis (Scardi *et al.*, 2018; Scardi, 2020; Leoni *et al.*, 2006; Yakovlev *et al.*, 2019; Leoni, 2014).

A key consequence of this formulation is that crystal shape and lattice symmetry are no longer rigidly coupled. This is particularly relevant for nanocrystals exhibiting anisotropic growth, strain-driven transformation, or shape-controlled synthesis, where external morphology is not necessarily aligned with crystallographic axes. Because many nanoscale systems exhibit non-crystallographic morphologies in which geometric and atomic ordering are strongly intertwined (Nguyen *et al.*, 2023; Xia *et al.*, 2009; Marks & Peng, 2016), the proposed framework enables direct extraction of shape–lattice orientation and deformation parameters from powder diffraction data without imposing crystallographic constraints on shape. We demonstrate the practical utility of the approach through analysis of shape-driven phase transformations in metal–alloy nanocrystals, where the relationship between initial and transformed states can be explicitly tracked. To our knowledge, this work presents a generalized affine-deformation extension of the CVF/WPPM formalism enabling independently refinable shape and lattice transformations.

## 2. Theory

### 2.1. Shape and lattice deformation configurations

To classify the deformation state of a given domain, which may involve all shape and lattice deformation and rotation, it is useful to encode the chosen combination of operations in a compact and unambiguous form. We therefore introduce a single integer label obtained from the binary representation of the applied deformation and rotation operations. To achieve this, we define a vector:

$$([SR],[SD],[LR],[LD]), \tag{1}$$

where $[SR]$, $[SD]$, $[LR]$, $[LD]$ are binary flags indicating whether the corresponding operation is applied. The letters denote shape (S) or lattice (L), and rotation (R) or deformation (D). For example, the vector representing a combined shape rotation $[SR]$ and lattice deformation $[LD]$ is $(1, 0, 0, 1)$. These four binary values can be interpreted as the bits of a 4-bit number; in this example, the sequence 1001 corresponds to the decimal value 9. In total, 16 combinations are possible, although only half are non-redundant. **Table 1** summarizes the complete set, noting that these relationships are independent of the specific lattice symmetry or the numerical values of the size and deformation parameters.

**Table 1** Combinations of shape, lattice rotations, and deformations. The binary vector $([SR],[SD],[LR],[LD])$ and its corresponding decimal representation are listed together with the equivalent transformation cases and an illustrative example of the resulting deformation/rotation relative to the undeformed reference state (case 0).

| Decimal Value | Binary Vector | Equivalent Vectors | Equivalent Values | Transformation Example |
|---|---|---|---|---|
| 0 | (0,0,0,0) | | | |
| 1 | (0,0,0,1) | | | |
| 2 | (0,0,1,0) | (1,0,0,0)<br>(1,0,1,0) | 8<br>10 | |
| 3 | (0,0,1,1) | (1,0,0,1)<br>(1,0,1,1) | 9<br>11 | |
| 4 | (0,1,0,0) | (1,1,0,0) | 12 | |
| 5 | (0,1,0,1) | (1,1,0,1) | 13 | |
| 6 | (0,1,1,0) | (1,1,1,0) | 14 | |
| 7 | (0,1,1,1) | (1,1,1,1) | 15 | |

### 2.1. Deformations of the crystal lattice and domain shape

The calculation of the PD profile from powders composed of monodisperse, monoatomic crystalline domains provides the reference framework for the generalized treatment developed in this work. The mathematical and crystallographic foundations underlying this formalism follow the established approaches commonly used in crystallography, total scattering, and line-profile (Coelho, 2018; Giacovazzo, 2011; Warren, 1990; Various, 2005; Krawitz, 2001; Scardi *et al.*, 2018; Scardi & Leoni, 2002; Leoni, 2014), and are summarized in appendices. **Appendix A** introduces the geometric framework of direct and reciprocal space, including the metric tensor (Eq. (A.1)), unit-cell volume (Eq. (A.2)), interplanar spacing relation (Eq. (A.3)), orthogonalization matrix (Eq. (A.4)), and reciprocal-space coordinates (Eq. (A.6)), together with their roles in defining distances, angles, scalar products, and reciprocal-space representations. **Appendix B** summarizes the affine transformation and deformation tensors used to describe lattice and domain distortions, together with their action on metric tensors and reciprocal vectors. **Appendix C** outlines the Whole Powder Pattern Modelling (WPPM) formalism and the Fourier treatment of diffraction from finite crystalline domains. These concepts establish the notation and mathematical framework adopted throughout the following derivations. To preserve analytical tractability and the validity of the Fourier-based treatment, the discussion is restricted to defect-free crystalline domains, or to systems containing only dilute defects for which lattice periodicity remains effectively preserved. **Table 2** summarizes the notation adopted throughout the following derivations. All deformation tensors are treated as passive affine transformations acting through pullback operations on the metric tensor and reciprocal-space coordinates. Together, these derivations establish a general framework for line-profile analysis of powder diffraction data from finite crystalline domains exhibiting shape and/or lattice deformations, and non-rigid shape–lattice orientation relationships.

**Table 2** **Summary of the symbols, tensors, coordinate frames, and metric relations**.

| Symbol | Meaning | Acts in / maps between |
|---|---|---|
| $\mathbf{A}$ | Direct lattice basis matrix | lattice basis → Euclidean frame |
| $\mathbf{M}$ | Orthogonalization matrix | crystallographic → orthonormal coordinates |
| $\mathbf{G}$ | Metric tensor | direct lattice space |
| $\mathbf{T}$ | Deformation tensor | domain/lattice transformation |
| $\boldsymbol{n}$ | Miller-index vector | |
| $\boldsymbol{n}_{\perp}$ | Orthogonalized reciprocal vector | |
| $\boldsymbol{q}$ | reciprocal-space scattering vector | |
| $V$ | Volume of the unit cell | direct lattice |
| $W$ | Volume of the crystal domain | real-space crystal domain |
| $d_n$ | Interplanar spacing | |

Note: Subscripts $s$, $d$, and $sd$ denote quantities associated with shape deformation, lattice deformation, and coupled shape–lattice deformation, respectively. $\mathbf{G}$ is symmetric positive-definite metric tensor.

The following subsections introduce the theoretical developments that extend the formalism summarized in the appendices to the deformation cases summarized in 2.1, and form the basis of all application examples presented in Section 3. Specifically, Sections 2.1.1, 2.1.2, and 2.1.3 derive how shape distortions, lattice deformations, rotations, and size dispersion can be incorporated within a unified framework. For each case, we show how the indicator function (and the associated common-volume function providing the shape Fourier coefficients), the lattice mapping, and the corresponding shape transformation operators are modified under the applied deformation, and how these modifications propagate consistently into the calculated diffraction profile. Throughout the following derivations, reciprocal-space vectors are represented in orthonormal coordinates using the transformation $\boldsymbol{n}_\perp = \mathbf{M}^{-1}\boldsymbol{n}$ of Eq. (A.6), while deformation tensors act as passive affine pullback operators on the metric tensor according to the formalism summarized in **Appendix B**.

#### 2.1.1. Domain shape deformation with fixed lattice

Shape deformation in the absence of lattice distortion can be efficiently introduced by applying a linear transformation to a known monoparametric reference shape through a diagonal deformation tensor **T**. This operation produces a domain whose shape is stretched, either elongated or contracted, along the principal axes while the underlying lattice remains unchanged (**Table 1** case 4, (0,1,0,0)). In this sense, the treatment parallels the results of Allegra and Wilson (Allegra & Wilson, 1983), but now it is formulated within a fully Fourier-based framework. Including a rotational component in the deformation tensor allows the deformed shape to be oriented arbitrarily with respect to the lattice basis, enabling, for example, the modelling of anisotropic growth along any crystallographic direction.

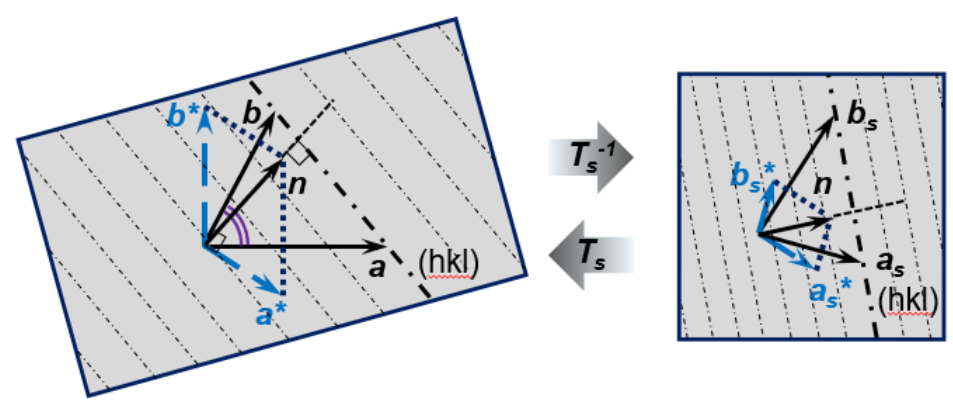

**Figure 1** Pure shape deformation: real (left) and auxiliary (right) spaces relation.

For practical computation, it is advantageous not to deform the shape function directly. Instead, the deformation is represented through a transformed virtual lattice in which the deformed domain is mapped onto the undeformed reference geometry (**Figure 1**). In this virtual-lattice space, the Fourier coefficients remain analytically available, while crystallographic quantities such as Bragg peak positions and oriented interatomic distances (Leonardi, 2021; Farrow *et al.*, 2007; Loopstra & Rietveld, 1969; Rietveld, 1967, 1969) remain referenced to the original undeformed lattice. A deformed crystallite domain $\widetilde{\Omega}$ is represented as the image of a reference domain $\Omega$ under an affine transformation tensor $\mathbf{T}_s$, such that $\widetilde{\omega} = \mathbf{T}_s\omega$. Rather than evaluating the Fourier coefficients directly on the deformed domain, the problem is pulled back onto the reference configuration through the inverse mapping $\mathbf{x} = \mathbf{T}_s^{-1}\tilde{\mathbf{x}}$. Under the affine transformation $\tilde{\mathbf{x}} = \mathbf{T}_s\mathbf{x}$, reciprocal-space vectors transform contravariantly in order to preserve the Fourier phase factor, $\mathbf{q}\cdot\mathbf{x} = \tilde{\mathbf{q}}\cdot\tilde{\mathbf{x}}$. Substituting $\tilde{\mathbf{x}} = \mathbf{T}_s\mathbf{x}$ gives $\tilde{\mathbf{q}} = \mathbf{T}_s^{-T}\mathbf{q}$. Accordingly, the deformation can be represented through the transformed metric tensor $\mathbf{G}_s = (\mathbf{T}_s^{-1})^T \cdot \mathbf{G} \cdot \mathbf{T}_s^{-1}$, while preserving the analytical expressions associated with the reference geometry. In this formulation, the diagonal elements of $\mathbf{T}_s$ encode differential stretching, whereas the rotational component defines the relative orientation between the principal axes of the deformed object and those of the lattice (see Eqs. (A.8) and (A.10)).

The deformation induces a modified metric for the virtual lattice,

$$\mathbf{G}_\mathrm{s} = (\mathbf{T}_\mathrm{s}^{-1})^\mathrm{T} \cdot \mathbf{A} \cdot \mathbf{A}^\mathrm{T} \cdot \mathbf{T}_\mathrm{s}^{-1} = \mathbf{M}_\mathrm{s} \cdot \mathbf{M}_\mathrm{s}^\mathrm{T}. \tag{2}$$

Using the metric-tensor formalism of Eq. (A.1), the deformation tensors are restricted to nonsingular affine transformations, ensuring that the transformed metric tensor $\mathbf{G}_\mathrm{s}$ remains symmetric positive definite. From this relation, an updated orthogonalization matrix $\mathbf{M}_\mathrm{s}$ can be obtained analogously to the Cholesky decomposition in Eq. (A.4). Recalling Eqs. (A.4), (A.5), and (A.6), in this virtual setting, the reciprocal basis and Miller index relations remain formally identical,

$$\boldsymbol{n}_{s,\perp} = \mathbf{M}_\mathrm{s}^{-1} \cdot \boldsymbol{n} \text{ and } |\boldsymbol{n}_{s,\perp}| = \det(\mathbf{M}_\mathrm{s}^{-1}) \cdot |\boldsymbol{n}| = \frac{|\boldsymbol{n}|}{V_s} = |\boldsymbol{n}| \det(\mathbf{T}_\mathrm{s}) \det(\mathbf{G})^{-\frac{1}{2}}, \tag{3}$$

but now refer to the deformed metric. The transformation $\mathbf{T}_\mathrm{s}$ also changes the volume of the associated unit cell,

$$\frac{V}{V_s} = \sqrt{\frac{\det(\mathbf{G})}{\det(\mathbf{G}_\mathrm{s})}} = \sqrt{\frac{\det(\mathbf{G})}{\det\left((\mathbf{T}_\mathrm{s}^{-1})^\mathrm{T} \cdot \mathbf{G} \cdot \mathbf{T}_\mathrm{s}^{-1}\right)}} = \sqrt{\frac{\det(\mathbf{G})}{\det\left((\mathbf{T}_\mathrm{s}^{-1})^\mathrm{T}\right) \det(\mathbf{G}) \det(\mathbf{T}_\mathrm{s}^{-1})}} = \frac{1}{\det(\mathbf{T}_\mathrm{s}^{-1})} = \det(\mathbf{T}_\mathrm{s}), \tag{4}$$

reflecting the deformation of the shape in real space.

Because the actual lattice is assumed to be unaltered, the original metric $\boldsymbol{G}$ is retained for crystallographic computations such as the evaluation of the Bragg peak positions (i.e., the interplanar spacing) according to Eq. (A.3) $d_{\boldsymbol{n}} = (\boldsymbol{n}^T \cdot \mathbf{G}^{-1} \cdot \boldsymbol{n})^{-\frac{1}{2}}$. The role of the virtual lattice is instead to provide the correct directions along which the Fourier coefficients of the shape transform must be evaluated. For a given direction $\boldsymbol{n}$, the corresponding direction in the virtual lattice is obtained using the transformed basis. The normalized Miller indices commonly used in Fourier descriptions of simple polyhedral shapes (e.g. in (Scardi & Leoni, 2001; Warren, 1990)) generalize here to

$$(A \quad B \quad C)^T = \frac{\boldsymbol{n}_{s,\perp}}{|\boldsymbol{n}_{s,\perp}|} = (\mathbf{M}_\mathrm{s} \cdot \boldsymbol{n}) \frac{V_s}{|\boldsymbol{n}|} = (\mathbf{M}_\mathrm{s} \cdot \boldsymbol{n}) \frac{V}{||\mathbf{T}_\mathrm{s}|| |\boldsymbol{n}|} = (\mathbf{M}_\mathrm{s} \cdot \boldsymbol{n}) \frac{\det(\mathbf{G})^{\frac{1}{2}}}{\det(\mathbf{T}_\mathrm{s}) |\boldsymbol{n}|}, \tag{5}$$

which ensures consistency between the deformed geometry and the Fourier representation. The cutoff value $K_{\tilde{\omega}}$ at which the Fourier coefficients vanish is likewise rescaled by the deformation,

$$\begin{aligned} K_{\tilde{\omega}} &= \frac{K_\omega}{k_s} \\ k_s &= \frac{d_{\boldsymbol{n}}}{d_{\boldsymbol{n}_s}} = \sqrt{\frac{\boldsymbol{n}_{s,\perp}^T \cdot \mathbf{G}_\mathrm{s}^{-1} \cdot \boldsymbol{n}_{s,\perp}}{\boldsymbol{n}^T \cdot \mathbf{G}^{-1} \cdot \boldsymbol{n}}} = \frac{|\mathbf{M}_\mathrm{s}^{-1} \cdot \boldsymbol{n}|}{|\mathbf{M}^{-1} \cdot \boldsymbol{n}|}, \end{aligned} \tag{6}$$

where $k_s$ captures the directional stretching introduced by $\mathbf{T}_\mathrm{s}$. With these adjusted quantities, the standard analytical expressions, proposed e.g., in (Scardi & Leoni, 2001), for the Fourier coefficients of simple convex shapes can be used directly.

Finally, the deformation $\mathbf{T}_\mathrm{s}$ modifies the volume of the finite crystal domain itself. The deformed domain volume is

$$W = W_s \frac{V}{V_s} = W_s \det(\mathbf{T}_s), \tag{7}$$

consistent with the geometric transformation applied to the shape. This mapping between the original and virtual lattices provides a unified mechanism for including shape deformation, together with arbitrary orientation, into the modelling of the powder-diffraction profile, while leaving the underlying lattice parameters and Bragg positions unchanged.

#### 2.1.2. Crystal lattice deformation with fixed domain shape and size

In certain situations, the primary interest lies not in deforming the crystallite shape, but in modelling modifications of the underlying lattice. Changes in lattice geometry can generate new reflections absent in the parent structure or eliminate others through symmetry enhancement. Such cases arise, for example, when modelling subtle distortions, phase transitions, or size/shape effects in materials with intrinsically low symmetry. In these situations, one often wishes to preserve the external domain shape and volume while transforming only the lattice.

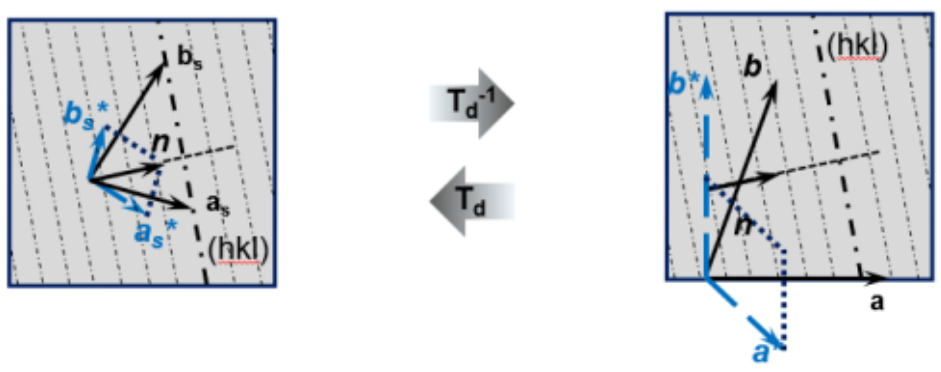

**Figure 2** Pure lattice deformation: real (left) and auxiliary (right) spaces relation.

Let $\mathbf{T}_\mathrm{d}$ denote the deformation applied to the lattice that maps the original structure onto the target one (**Figure 2**). As an illustrative example, $\mathbf{T}_\mathrm{d}$ may transform a face-centred cubic lattice into a face-centred tetragonal one. The metric of the deformed lattice follows directly from the transformation:

$$\mathbf{G}_\mathrm{d} = \left(\mathbf{T}_\mathrm{d}^{-1}\right)^\mathrm{T} \cdot \mathbf{G} \cdot \mathbf{T}_\mathrm{d}^{-1} = \left(\mathbf{T}_\mathrm{d}^{-1}\right)^\mathrm{T} \cdot \mathbf{A} \cdot \mathbf{A}^\mathrm{T} \cdot \mathbf{T}_\mathrm{d}^{-1} = \mathbf{M}_\mathrm{d} \cdot \mathbf{M}_\mathrm{d}^\mathrm{T}, \tag{8}$$

where $\mathbf{A}$ is the orthogonalization matrix of the original lattice and $\mathbf{M}_\mathrm{d}$ is the corresponding matrix for the deformed lattice. This metric must now be used consistently for all crystallographic calculations, most notably for coordinate transformations into orthonormal space,

$$\begin{aligned} \boldsymbol{n}_{d,\perp} &= \mathbf{M}_\mathrm{d}^{-1} \cdot \boldsymbol{n} \\ \left|\boldsymbol{n}_{d,\perp}\right| &= \det\left(\mathbf{M}_\mathrm{d}^{-1}\right) \cdot |\boldsymbol{n}| = \frac{|\boldsymbol{n}|}{V_d} = |\boldsymbol{n}| \det(\mathbf{T}_\mathrm{d}) \det(\mathbf{G})^{-\frac{1}{2}}, \end{aligned} \tag{9}$$

and for the evaluation of interplanar spacings, still remain valid Eq. (A.3) $d_{\boldsymbol{n}} = \left(\boldsymbol{n}^T \cdot \mathbf{G}_d^{-1} \cdot \boldsymbol{n}\right)^{-\frac{1}{2}}$, with $\mathbf{G}_{\boldsymbol{d}} = \mathbf{G}$. The same transformation governs the direction along which the Fourier coefficients of the shape transform must be evaluated:

$$(A \quad B \quad C)^T = \frac{\boldsymbol{n}_{d,\perp}}{\left|\boldsymbol{n}_{d,\perp}\right|} = \left(\mathbf{M}_\mathrm{d}^{-1} \cdot \boldsymbol{n}\right) \frac{V_d}{|\boldsymbol{n}|} = \left(\mathbf{M}_\mathrm{d}^{-1} \cdot \boldsymbol{n}\right) \frac{\det(\mathbf{G})^{\frac{1}{2}}}{\det(\mathbf{T}_\mathrm{d})|\boldsymbol{n}|}.$$

(10)

This ensures consistency between the deformed lattice basis and the Fourier-space peak-profile formalism.

Importantly, because the deformation affects only the lattice and not the crystal domain shape, neither the domain volume nor the cutoff position of the Fourier coefficients is altered relative to the standard expressions for the chosen shape. The Fourier treatment of size broadening therefore proceeds exactly as in the undeformed case, while the lattice deformation manifests through the modified metric, altered reciprocal-lattice geometry, and resulting changes in peak positions and potential symmetry-induced selection rules.

#### 2.1.3. Coupled domain shape and crystal lattice deformations

In many practical situations, the deformation affecting a crystalline domain cannot be attributed solely to a change in shape or solely to a change in lattice geometry; rather, both effects occur simultaneously. Coupling shape and lattice deformation enables the modelling of complex crystallite morphologies embedded within lattices of reduced symmetry, starting from a known reference shape defined in a higher-symmetry lattice. To account for this general condition, we distinguish the lattice deformation tensor $\mathbf{T}_\mathrm{d}$ from the shape deformation tensor $\mathbf{T}_\mathrm{s}$, each acting on different components of the structural model.

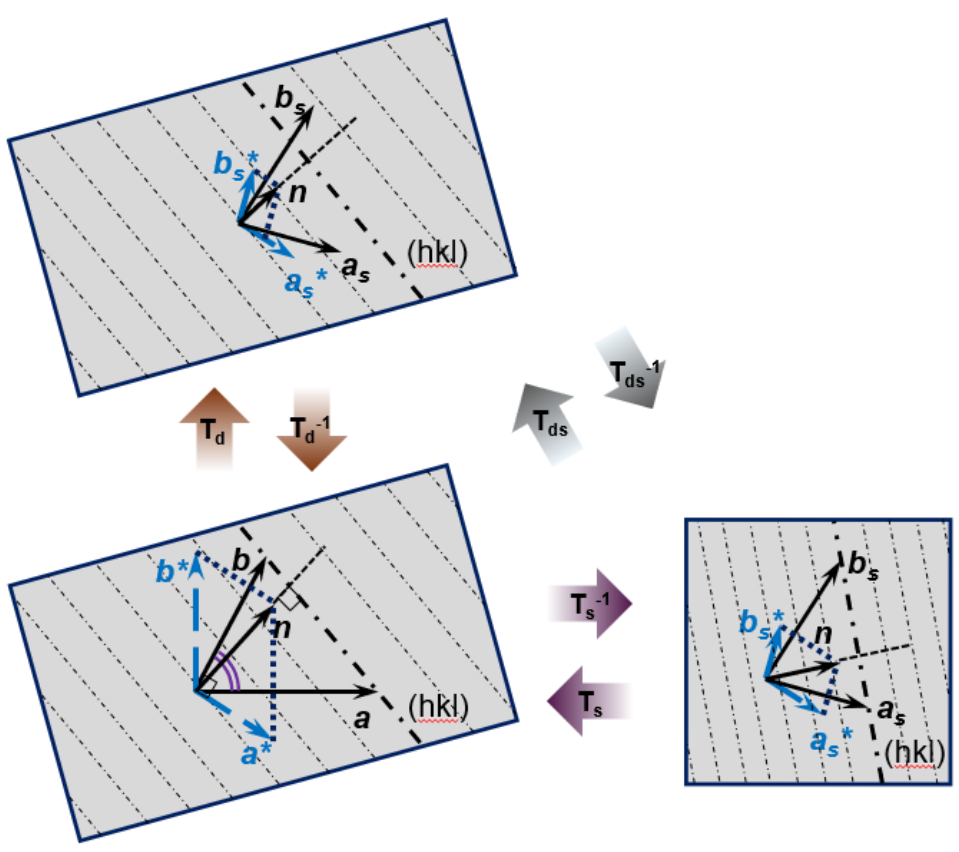

**Figure 3** Couple shape-lattice deformation: real (left-top) and auxiliary (right-bottom) spaces relation through the intermediate pure lattice deformation (left-bottom).

The lattice deformation modifies the metric of the underlying Bravais lattice according to

$$\mathbf{G}_\mathrm{d} = \left(\mathbf{T}_\mathrm{d}^{-1}\right)^\mathrm{T} \cdot \mathbf{G} \cdot \mathbf{T}_\mathrm{d}^{-1}, \quad (11)$$

while the shape deformation must be applied on top of this already transformed lattice. The combined transformation tensor is obtained through sequential composition of the corresponding pullback mappings (**Figure 3**). The combined effect of both transformations is therefore described by the composite tensor $\mathbf{T}_{\mathrm{ds}} = \mathbf{T}_{\mathrm{d}} \cdot \mathbf{T}_{\mathrm{s}}$, and the metric experienced by the deformed domain becomes (according with **Appendix B**):

$$\mathbf{G}_{\mathrm{ds}} = (\mathbf{T}_{\mathrm{s}}^{-1})^{\mathrm{T}} \cdot \mathbf{G}_{\mathrm{d}} \cdot \mathbf{T}_{\mathrm{s}}^{-1} = \left(\mathbf{T}_{\mathrm{ds}}^{-1}\right)^{\mathrm{T}} \cdot \mathbf{G} \cdot \mathbf{T}_{\mathrm{ds}}^{-1} = \mathbf{M}_{\mathrm{ds}} \cdot \mathbf{M}_{\mathrm{ds}}^{\mathrm{T}}. \quad (12)$$

This expression makes clear that both deformations must be accounted for when determining the geometric and reciprocal-space properties of the transformed crystal.

For crystallographic quantities such as Bragg peak positions or interplanar spacings, only the lattice metric is required, and these are obtained as:

$$d_{\boldsymbol{n}} = \left(\boldsymbol{n}^T \cdot \mathbf{G}_{\mathrm{ds}}^{-1} \cdot \boldsymbol{n}\right)^{-\frac{1}{2}}. \quad (13)$$

However, for evaluating the Fourier coefficients associated with the finite crystallite shape, needed for computing peak profiles, the combined deformation must be considered explicitly. The direction along which the Fourier coefficients must be computed becomes:

$$(A \quad B \quad C)^T = \frac{\boldsymbol{n}_{ds,\perp}}{|\boldsymbol{n}_{ds,\perp}|} = \left(\mathbf{M}_{\mathrm{ds}}^{-1} \cdot \boldsymbol{n}\right)\frac{V_{ds}}{|\boldsymbol{n}|} = \left(\mathbf{M}_{\mathrm{ds}}^{-1} \cdot \boldsymbol{n}\right)\frac{\det(\mathrm{G})^{\frac{1}{2}}}{\det(\mathbf{T}_{\mathrm{ds}})|\boldsymbol{n}|}, \quad (14)$$

where $\boldsymbol{n}_{ds,\perp} = M_{ds}^{-1}\boldsymbol{n}$ is the direction expressed in the orthonormal frame of the doubly-deformed lattice and $V_{ds}$ is the deformed unit-cell volume.

Coupling shape and lattice deformation also affects both the unit-cell volume and the volume of the crystallite. The transformed unit-cell volume is:

$$\frac{V}{V_{ds}} = \sqrt{\frac{\det(\mathbf{G})}{\det(\mathbf{G}_{\mathrm{ds}})}} = \sqrt{\frac{\det(\mathbf{G})}{\det\left(\left(\mathbf{T}_{\mathrm{ds}}^{-1}\right)^{\mathrm{T}} \cdot \mathbf{G} \cdot \mathbf{T}_{\mathrm{ds}}^{-1}\right)}} = \sqrt{\frac{\det(\mathbf{G})}{\det\left(\left(\mathbf{T}_{\mathrm{ds}}^{-1}\right)^{\mathrm{T}}\right)\det(\mathbf{G})\det\left(\mathbf{T}_{\mathrm{ds}}^{-1}\right)}} = \frac{1}{\det\left(\mathbf{T}_{\mathrm{ds}}^{-1}\right)} = \det(\mathbf{T}_{\mathrm{ds}}), \quad (15)$$

and the volume of the deformed domain follows as:

$$W = W_{ds}\frac{V}{V_{ds}} = W_{ds}\det(\mathbf{T}_{\mathrm{ds}}). \quad (16)$$

Finally, the cutoff value at which the Fourier coefficients vanish, determined by the crystallite size along the chosen direction, must be rescaled to reflect the combined deformation. The scaling factor is:

$$k_{ds} = \frac{d_{\boldsymbol{n}}}{d_{\boldsymbol{n}_{ds}}} = \sqrt{\frac{\boldsymbol{n}_{ds,\perp}^T \cdot \mathbf{G}_{\mathrm{ds}}^{-1} \cdot \boldsymbol{n}_{ds,\perp}}{\boldsymbol{n}^T \cdot \mathbf{G}^{-1} \cdot \boldsymbol{n}}} = \frac{|\mathbf{M}_{\mathrm{ds}}^{-1} \cdot \boldsymbol{n}|}{|\mathbf{M}^{-1} \cdot \boldsymbol{n}|}, \quad (17)$$

ensuring that the shape transform is evaluated consistently with the geometry of the doubly-deformed crystal. Together, these relations provide a complete and unified description of systems in which both lattice and shape deformations contribute to the diffraction signal, covering the most general case encountered in realistic nanostructured materials.

### 2.1.4. Powder size distribution

A compact and efficient strategy for incorporating crystal size distributions into the diffraction model consists of averaging the Fourier coefficients directly over the size distribution, rather than integrating the full diffraction profiles [21]. Following this approach, the Fourier coefficient for a direction $\hat{\boldsymbol{v}}$ is computed as a volume-weighted average of the size-dependent coefficients $A_\omega(\hat{\mathbf{v}}, L, D)$:

$$A(\hat{\boldsymbol{v}}, L) = \frac{\int_{LK_{\omega,\hat{\boldsymbol{v}}}}^{\infty} A_\omega(\hat{\boldsymbol{v}}, L, D) V_\omega(D) p(D) \partial D}{\int_0^\infty V_\omega(D) p(D) \partial D}, \tag{18}$$

where $A_\omega(\hat{\mathbf{v}}, L, D)$corresponds to the directional Fourier coefficients derived from the finite-domain formalism summarized in **Appendix C**, $p(D)$ is the size distribution, and $V_\omega(D)$ the volume of a domain of characteristic size $D$. This formulation greatly reduces computational complexity because the integration is performed directly on the analytical expressions of the Fourier coefficients, bypassing the need to compute and average full diffraction patterns. Closed-form solutions of Eq. (18) exist for the most common probability distributions and domain shapes [18], [21], [50].

The same strategy can be extended to deformed domains. However, a limitation arises because the integral in Eq. (18) applies naturally only to shapes described by a single size parameter. In the present framework, deformation generally produces crystallites characterized by three independent size parameters as well as a specific orientation of their principal axes with respect to the lattice vectors. Analytical averaging is still possible in certain cases, such as when one or more size ratios are fixed, but in the most general scenario, numerical integration is required. Despite this, the approach remains computationally efficient and provides a practical route for incorporating realistic size distributions into diffraction-profile modelling, even in the presence of complex coupled deformations.

### 2.2. Integrating shape and lattice deformation in reciprocal and real-space total scattering analysis

The formalism developed above can be incorporated directly into the WPPM framework, where the implementation is particularly straightforward. The expressions that we derived extend those already used in WPPM to compute peak profiles. Consistent with the Fourier-domain intensity formalism introduced in Eq. (A.14), the contribution to the diffraction signal associated with a direction $\hat{\boldsymbol{v}}$ is written as:

$$I(\hat{\boldsymbol{v}}, Q) = \int_{-\infty}^{\infty} [A(\hat{\boldsymbol{v}}, L, D_\omega) + i\, B(\hat{\boldsymbol{v}}, L, D_\omega)]\, e^{iLQ}\, dL, \tag{19}$$

where the Fourier coefficients $A$ and $B$ incorporate finite-size effects, domain shape, and the combined shape–lattice deformation tensors. The full powder pattern is then obtained by summing all relevant $\hat{\boldsymbol{v}}$ directions, restricted in practice to those generating observable reflection peaks, with peak intensities either refined independently (e.g., Powley, Le Bail methods (Le Bail *et al.*, 1988; Pawley, 1981)) or constrained by a structural model (e.g., Rietveld method (Rietveld, 1967)). In this way, the deformation tensors, virtual-lattice transformation, and directional Fourier coefficients integrate seamlessly into WPPM without altering its core formulation.

Although WPPM provides the most direct reciprocal-space implementation, the same mathematical ingredients can be used to model real-space total-scattering signals. For the test cases presented hereinafter, we use the WPDFM method, which is the real-space counterpart of WPPM and provides a practical implementation pathway that naturally benefits from lattice-deformation mappings and the identification of independent crystallographic directions in base lattices such as face-centred, body-centred, and simple cubic. In this context, deformation tensors rescale or rotate the pair-distance grid used to evaluate the common-volume function $F_{\omega}(\hat{\boldsymbol{v}}, L)$ or, more generally, the directional Fourier coefficients (Eqs. (A.12) and (A.13)). The corresponding expression for the intensity becomes

$$I(Q) \propto f^2 \sum_n \sum_r F_{\omega}\,(\hat{\boldsymbol{v}}, L)\, \frac{\sin{(LQ)}}{LQ}, \tag{20}$$

which is adapted from the Debye scattering equation (Leonardi & Bish, 2016; Debye, 1913). Importantly, incorporating shape and lattice deformation directly within the WPDFM framework ensures that all diffuse-scattering contributions, most notably the thermal diffuse scattering (Urban, 1975), are modified in a manner fully consistent with the deformed domain geometry.

The same formalism applies equally well to other real-space PDF-fitting strategies, including methods such as the Interpolation Method (Yurchenko *et al.*, 2016, 2015; Yurchenko, 2014). Within such approaches, the deformation tensors introduced here play the same role: they modify the sampling of pair distances and the orientation of crystallographically independent directions, while the underlying shape transform governs the decay and modulation of the PDF. Thus, whether implemented within a WPDFM-style workflow or a PDF-interpolation framework, the mathematical treatment of size, shape, and lattice deformation is formally equivalent to that used in WPPM (Leonardi, 2021; Stokes & Wilson, 1942; Wilson, 1955). Whether applied in reciprocal space through WPPM or in real space through WPDFM or PDF-Interpolation methods, the combined use of deformation tensors, virtual-lattice transformations, and direction-dependent Fourier coefficients provides a unified and flexible strategy for modelling diffraction and total-scattering signals from domains exhibiting coupled shape and lattice distortions.

### 3. Results

To demonstrate the capabilities of the proposed formalism and to evaluate the accuracy of the modelling approach, virtual powder X-ray diffraction datasets were first generated using the Debye scattering equation (DSE) (Leonardi & Bish, 2016). These Debye-generated PD profiles are based on fully atomistic models with known shape and lattice deformation and therefore serve as surrogate experimental datasets. The resulting profiles were then analysed using the WPDFM framework extended by the theoretical derivations introduced in Section 2.1, which enables line-profile analysis accounting for coupled shape and lattice deformation. While the WPPM framework offers a direct reciprocal-space route, the WPDFM method is particularly advantageous for the present test cases because it handles detailed shape effects and is naturally compatible with the DSE formalism across both wide- and small-angle regimes. Within this framework, the deformation parameters were refined by simulated annealing, used solely as a numerical optimisation strategy. The refined deformation parameters obtained from the analysis were finally compared with the nominal values used to construct the atomistic models, providing a direct validation of the proposed approach.

All benchmark simulations were performed using the fcc Pd structure (unit-cell parameter 0.389070 nm). For each test case, an ensemble of 100 nanocrystals with the desired shape, size, and orientation was generated, with the lattice origin randomly placed within the unit cell to mimic the translational disorder expected in real powders (Minami & Ino, 1979). The powder-diffraction profiles were obtained by summing the individual DSE intensities of all nanocrystals in the ensemble. This procedure was applied to all simulated Pd test cases discussed below. The only exception is the final, most realistic example addressing the Cu–Pd solid-state phase transformation, where atomistic configurations were generated from molecular-dynamics simulations and the diffraction signal was computed directly from the MD snapshots (Leonardi *et al.*, 2011; Brandstetter *et al.*, 2008).

### 3.1. Shape elongation

Spherical Pd nanocrystals with a 10 nm diameter were used as a root reference. Ellipsoids were generated by elongating the sphere along the [100] direction from 10 nm to 60 nm in increments of 5 nm (0–600% elongation). Representative diffraction profiles optimized by WPDFM modelling for the 0%, 250%, and 500% elongation cases are shown in **Figure 4A**. Model refinement was carried out using simulated annealing, starting from parameters corresponding to an 8 nm Pd sphere. The WPDFM-based line-profile analysis of the Debye-simulated PD profiles follows the deformation state $(0,1,0,0)$ (i.e., case 4 in Table 1) and implements the "*domain shape deformation with fixed lattice*" formalism introduced in Section 2.1.1, specifically using Eqs. (5), (6), and (7).

As expected, increasing elongation modifies the decay envelope of the PDF, shifting its maximum toward shorter pair distances (**Figure 4B**). This behaviour reflects the intrinsic anisotropy introduced by shape deformation and highlights the inadequacy of assuming spherical particles when analysing PDF data of anisotropic nanoparticles. The WPDFM approach recovers the size and aspect ratio of the ellipsoids with negligible error relative to the DSE-generated profiles; residual differences diminish with increasing scattering momentum $Q$, due to the decay of the Pd atomic scattering factor. The refined ellipsoid axes and corresponding deformations closely reproduce the nominal values (**Figure 4C**). Minor discrepancies originate from the smooth approximation of the atomistic shape through the common-volume function. Equivalent accuracy is obtained for ellipsoids elongated along arbitrary orientations within the first Brillouin zone.

We further extended the deformation state of case 4 (Table 1) from (0,1,0,0) to (1,1,0,0), thereby completing the coverage of all cases belonging to group 4. A set of ellipsoids deformed by 25% and rotated was simulated along the conventional high-symmetry $\Gamma$-$K$-$M$-$\Gamma$ path, following the standard crystallographic convention (Setyawan & Curtarolo, 2010), with Γ, K, and M corresponding to the [100], [110], and [111] directions, respectively. The refined sizes and directional parameters (**Figure 4D**) exhibit excellent agreement, with orientation errors remaining below 1% in the most challenging regions between [110] and [111], where shape-truncated pair distances vary only subtly across directions.

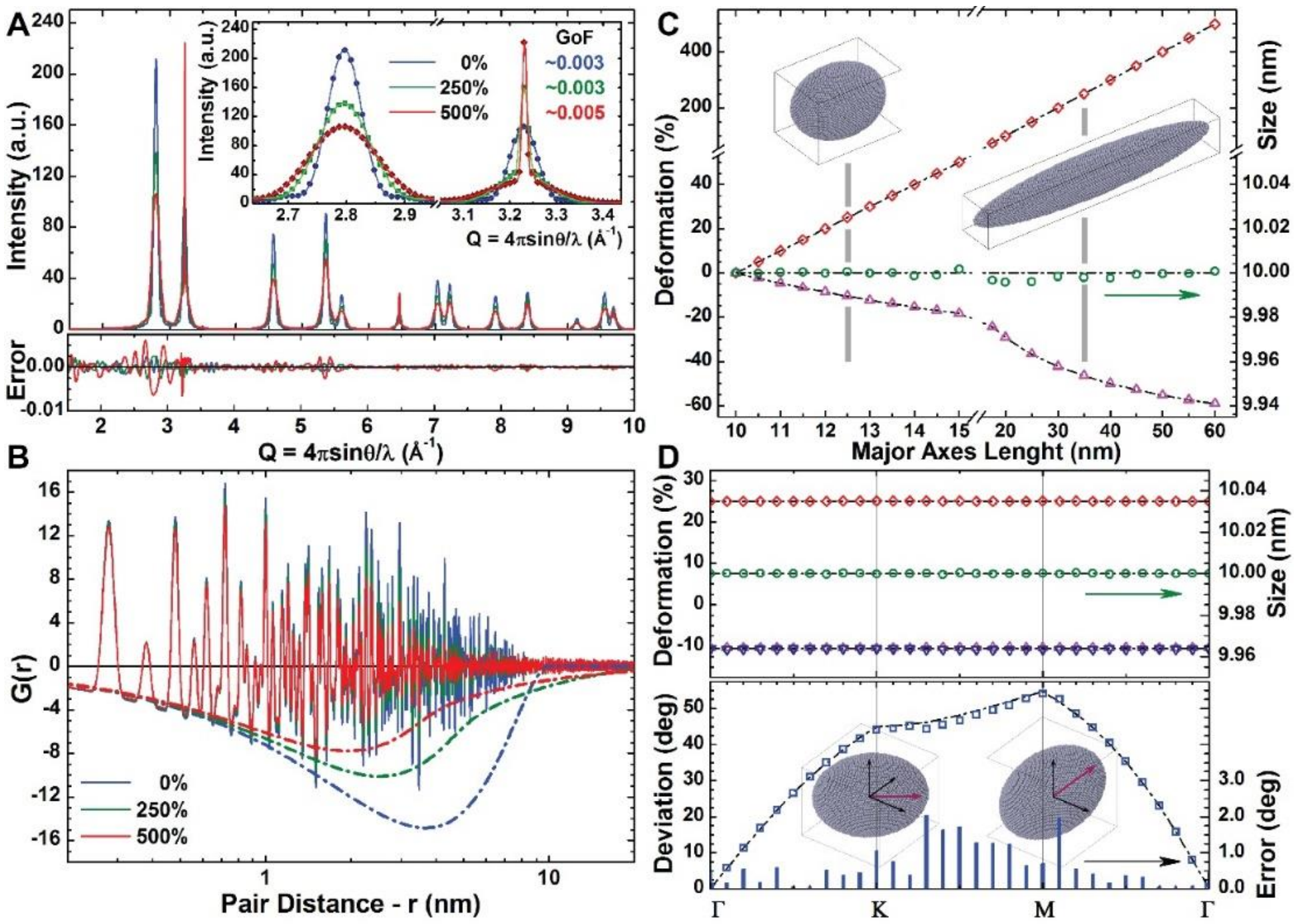

**Figure 4** Shape deformation: deformation states **A-C** (0,1,0,0) and **D** (1,1,0,0). Modelling of powder scattering data from powders of fcc Pd ellipsoidal nanocrystals with constant volume (unit cell 0.389070 nm). **A**, Comparison between DSE simulation (dots) and WPDFM modelling (line) and **B**, reduced pair distribution function (continuous line) and small-angle shape function corrections (dot-dash line) for the ellipsoids elongated along [100] by 0%, 250% and 500%. **C**, Ellipsoid axes and corresponding deformation obtained via WPDFM modelling of the profiles (dot) are compared with expected values (black line). **D**, Size, longitudinal and transversal deformations, and deviation of the axes of elongation from the [100] direction for ellipsoidal crystals in the first Brillouin zone.

### 3.2. Lattice and shape relative misorientation

Cylindrical nanocrystals present additional modelling challenges due to the coexistence of curved surfaces, sharp edges, and flat facets. Powder diffraction patterns were simulated for monodisperse cylinders (base diameter = 10 nm, height = 10 nm) with their symmetry axis oriented along various directions relative to the *fcc* base cartesian system within the Brillouin zone. Representative DSE profiles and corresponding WPDFM fits for cylinders aligned along [100], [110], and [111] are shown in **Figure 5A**. The WPDFM-based line-profile analysis of the Debye-simulated PD profiles follows the deformation state $(0,0,1,0)$ (i.e., case 2 in Table 1) and implements the "*Crystal lattice deformation with fixed domain shape and size*" formalism introduced in Section 2.1.2, specifically using Eq. (9).

Among the reflections, the [111] peak is most sensitive to the orientation of the cylinder axis, and accordingly, this case shows the largest modelling deviations (**Figure 5)**. Nonetheless, parameter extraction remains robust, with size and orientation retrieved with negligible error. The angular dependence of model discrepancies mirrors that observed for ellipsoids, indicating that small variations in shape-truncated pair distances near the [110]–[111] manifold limit the discriminability of closely related orientations.

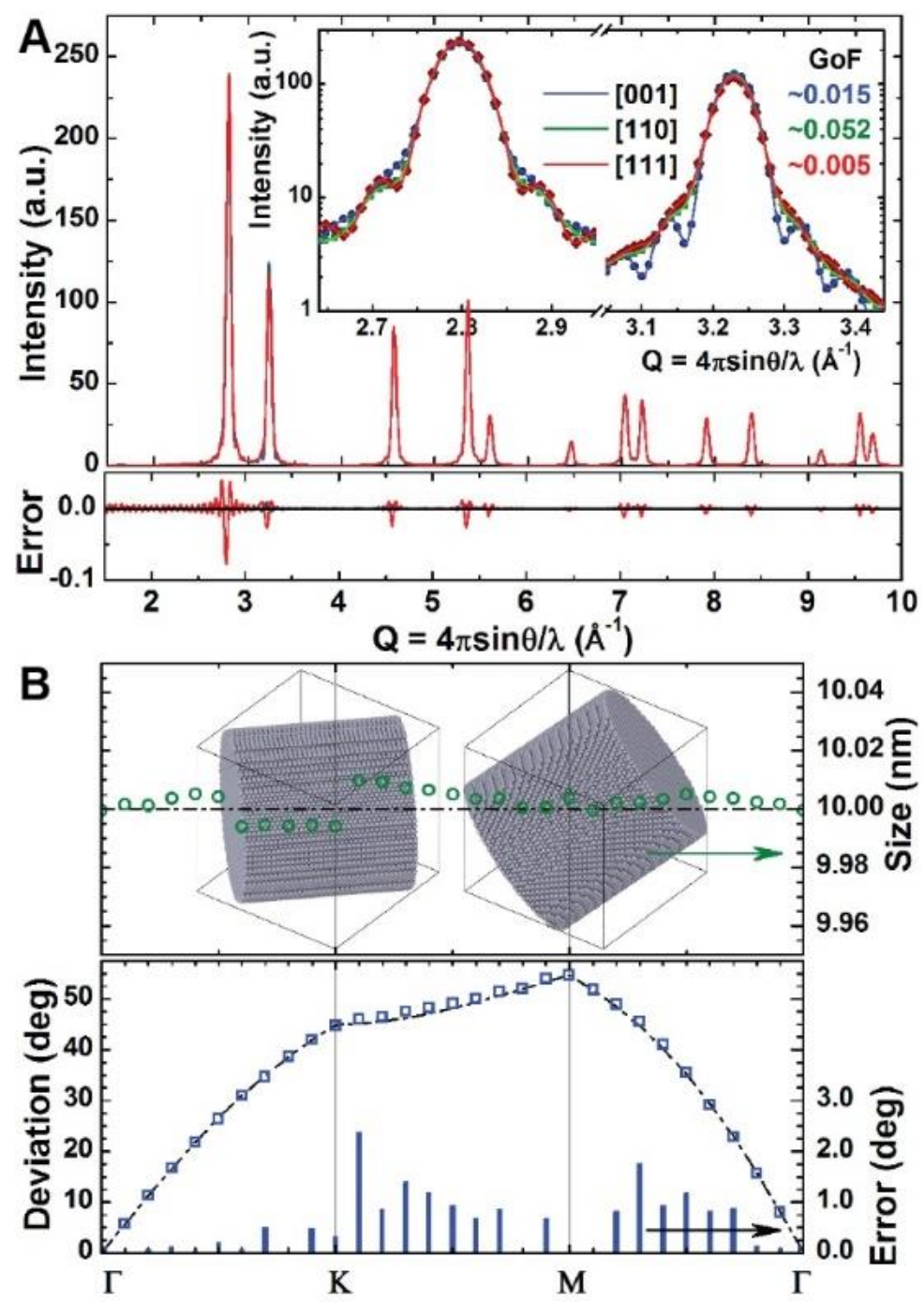

**Figure 5** Shape rotation relative to the lattice base: deformation state $(0,0,1,0)$. Modelling of powder scattering data from powders of fcc Pd cylindrical nanocrystals with height equal to the base diameter (unit cell 0.389070 nm and size 10nm). **A**, scattering intensity profile from crystals with the cylinder axes along the [100], [110], and [111] crystallographic directions. The profiles modelled using the WPDFM method and shape rotation matrix (line) are compared with the average solution of the Debye scattering equation for a set of 100 equivalent crystals with a random position of the cylinder centre, relative to the lattice origin (dot). **B**, size and deviation of the cylinder axes from the [100] direction estimated via global optimization of the agreement between Debye simulated and WPDFM modelled profiles (dot) are compared with nominal values (black line) for crystals with axes of the cylindrical shape varying along the $\overline{\Gamma K M \Gamma}$path ($\Gamma = [100]$, $K = [110]$, $M = [111]$) in the first Brillouin zone.

### 3.3. Shape deformation and misorientation relative to the lattice

Unlike ellipsoids and cylinders, hexagonal prisms in an *fcc* lattice have sharp edges and flat facets that introduce more complex, orientation-dependent peak shapes. Existing analytical shape functions cover only a limited subset of such geometries and do not explicitly incorporate orientation. To test the generality of the proposed framework, three families of nanoprisms were generated: (i) prisms with edge $e = 16$ nm and height $h = 16$ nm oriented along [001], (ii) platelet-like prisms with $h/e = 1/4$ oriented along [110], and (iii) rod-like prisms with $h/e = 5$ oriented along [111]. The WPDFM-based line-profile analysis of the Debye-simulated PD profiles follows the deformation state $(1,0,1,0)$ (i.e., case 6 in Table 1) and implements the "*coupled domain shape and crystal lattice deformations*" formalism introduced in Section 2.1.3, specifically using Eqs. (13), (14), (15), (16), and (17).

**Figure 6** shows the particle shapes, the corresponding DSE powder patterns, and the WPDFM modelling results. Differences in peak shapes, particularly the pointed peaks arising from sharp prism edges, are accurately reproduced. As expected, differences in the PDF at long distances (**Figure 6C**) reflect the varying effective lengths of the prisms along different orientations. In real datasets, such shape-dependent effects are further moderated by noise and background contributions, which tend to homogenize residuals across peaks.

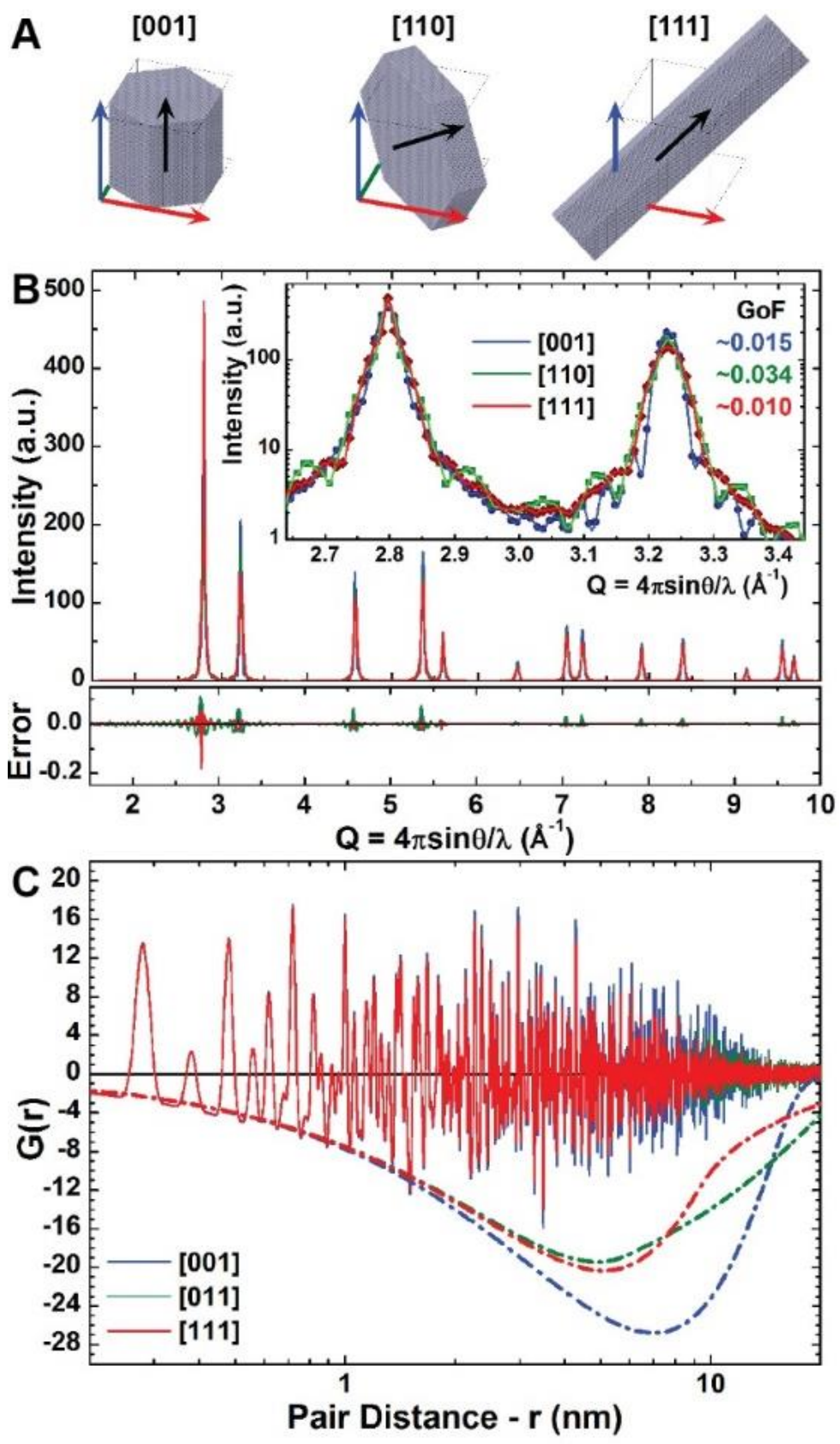

**Figure 6** Combined deformation and rotation of the shape: deformation state $(1,0,1,0)$. Modelling of powder scattering data from powders of fcc Pd nanocrystals with hexagonal prism shape and different longitudinal-to-transversal size shape ratio (unit cell 0.389070 nm, size 16nm). **A**, nanoprism with axes along the [100], [110], and [111] crystallographic directions with a size ratio 1:4 for the [110]

platelet-like, and size ratio 5:1 for the [111] rod-like shapes. **B**, the scattering intensity profile modelled using the WPDFM method and shape rotation matrix (line) are compared with the average solution of the Debye scattering equation for a set of 100 equivalent crystals with random position of the shape relative to the lattice structure (dot). **C**, pair-distribution-function (continuous line) and small-angle shape function corrections (dot-dash line).

### 3.4. Lattice deformation

To demonstrate the impact of pure lattice deformation, primitive-cubic (P) Pd nanocrystals were transformed into base-centred (C), body-centred (I), or face-centred (F) lattices following the deformation paths illustrated in **Figure 7A**. The diffraction patterns of the fully transformed lattices (**Figure 7B**) exhibit expected symmetry-driven changes: some reflections disappear due to systematic absence conditions, while others change intensity because of altered multiplicities. The corresponding PDFs (**Figure 7C**) show identical interatomic distances for base-centred and face-centred lattices but in different multiplicities, as expected from the number of centred faces. Both the intensity PD and PDF profiles were computed with the WPDFM method following the "*crystal lattice deformation with fixed domain shape and size*" formalism introduced in Section 2.1.2, specifically using Eq. (10) to describe the deformation state $(0,0,0,1)$ (i.e., case 1 in Table 1) .

The transformation from a primitive (P) cell to the base-centred (C), body-centred (I), and face-centred (F) representations associated with the same cubic lattice (i.e., describing the same underlying cubic Bravais lattice with identical lattice parameters) can be described as a linear transformation of the basis vectors. By parameterising the lattice deformation through a transformation fraction $l \in [0,1]$, a continuous sequence of intermediate lattice states can be generated (**Figure 7D–F**). This transformation involves both a rescaling of the basis vector lengths and a corresponding change in the interaxial angles. For example, the conversion from the primitive to the face-centred setting transforms the lattice parameters from $a = 3.89070$ Å with 90° interaxial angles to $a = 2.75114$ Å with 60° angles. The transformation can be expressed as a one-parameter family of linear mappings $\mathbf{T}_{\mathrm{d}}(l) = \mathbf{I} + l(\mathbf{T}_{end} - \mathbf{I})$, $l \in [0,1]$, where $\mathbf{T}_{end}$ is the basis transformation matrix corresponding to the target centred lattice. This formulation provides a continuous interpolation between the primitive cubic basis ($l = 0$) and the primitive basis of the centred lattice ($l = 1$). The end-member transformations are given by,

$$\mathbf{T}_{\mathrm{C}} = \begin{bmatrix} 1/2 & 1/2 & 0 \\ -1/2 & 1/2 & 0 \\ 0 & 0 & 1 \end{bmatrix}, \mathbf{T}_{\mathrm{I}} = \frac{1}{2}\begin{bmatrix} -1 & 1 & 1 \\ 1 & -1 & 1 \\ 1 & 1 & -1 \end{bmatrix}, \text{and } \mathbf{T}_{\mathrm{F}} = \frac{1}{2}\begin{bmatrix} 0 & 1 & 1 \\ 1 & 0 & 1 \\ 1 & 1 & 0 \end{bmatrix}, \quad (21)$$

for base-centred ($\mathbf{T}_{\mathrm{C}}$), body-centred ($\mathbf{T}_{\mathrm{I}}$), and face-centred ($\mathbf{T}_{\mathrm{F}}$), respectively.

The corresponding PDFs (**Figure 7G–I**) show a smooth evolution of interatomic separations and intensities, demonstrating that the deformation tensor framework naturally bridges symmetry-related structures along a continuous transformation path.

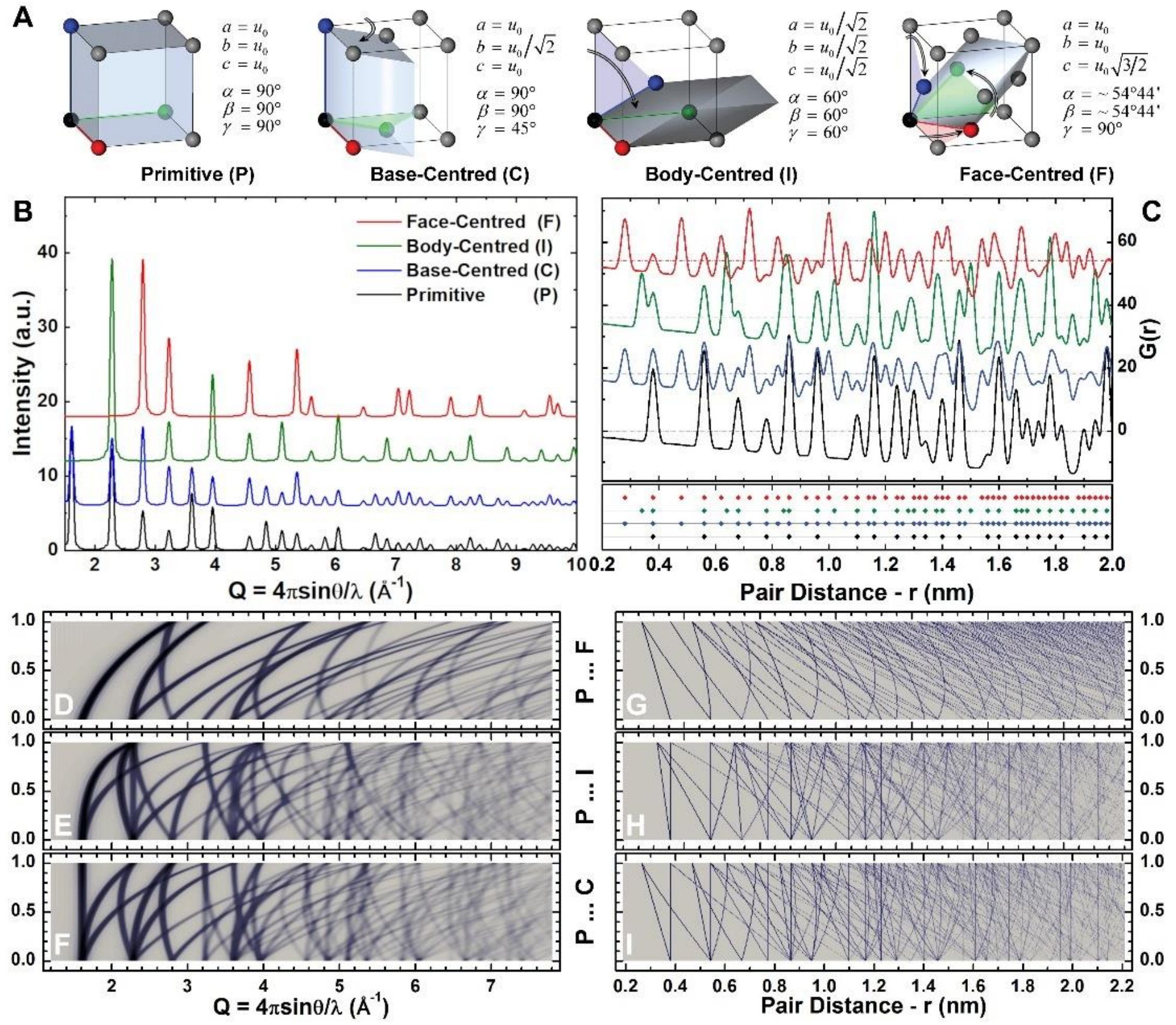

**Figure 7** Deformation of the lattice: deformation state (0,0,0,1). Modelling of powder scattering data from powders of Pd spherical nanocrystals with different lattice structures (size 10nm). **A**, base-, body-, and face-centered end-member structures obtained from the deformation of a cubic primitive unit cell (cell parameter 0.389070nm). **B** and **C**, intensity profile and pair-distribution-function modelled for powders of sphere crystals with the end-member structures outlined in **A**. **D** to **F** and **G** to **I**, respectively, intensity and pair-distance profiles evolution with crystal structure varying between primitive (bottom) and deformation end-members (top) according to the transformation paths drawn in **A**.

### 3.5. Simultaneous shape and lattice deformation

Simultaneous deformation of the lattice and particle shape was tested by modelling the Bain path linking *fcc* Pd to its metastable $bct_{10}$-like configuration (**Figure 8A**). Along this path, the reduction of the $c/a$ ratio modifies both the lattice metric and the crystallite morphology: spheres transform into ellipsoids, cubes into parallelepipeds, and so on. The WPDFM-based line-profile analysis of the Debye-simulated PD profiles follows the deformation state (0,1,0,1) (i.e., case 5 in Table 1) and implements the "*coupled domain shape and crystal lattice deformations*" formalism introduced in Section 2.1.3, specifically using Eqs. (13), (14), (15), (16), and (17).

Using only lattice deformation (**Figure 8B**) captures the first-order peak-position changes but fails to reproduce subtle modulations, such as the intensity ripples associated with finite crystal dimensions. Incorporating both lattice and shape deformation (**Figure 8C**) yields markedly improved agreement, especially in these ripple regions. The corresponding difference plots become nearly featureless, emphasizing the necessity of including both deformation modes to model diffraction profiles accurately. This combined-tensor strategy generalizes to any solid-state transformation, linking peak-shape evolution directly to the underlying changes in lattice geometry.

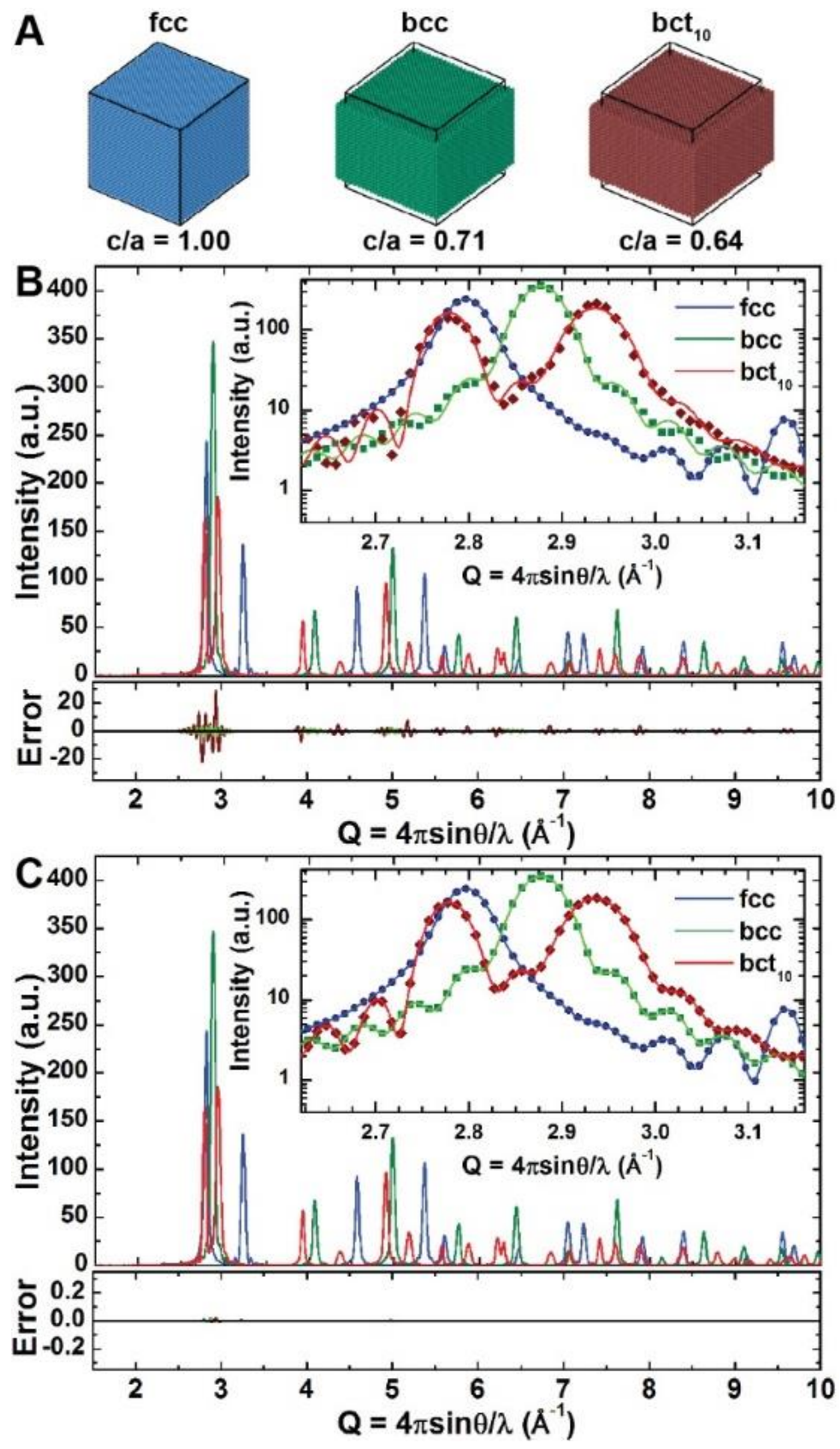

**Figure 8** Coupled lattice and shape deformations. Modelling of powder scattering data from powders of Pd nanocrystals along the Bain deformation path (unit cell 0.389070 nm, size 10nm) (Edgar C. Bain, 1924). **A**, fcc nanocube deforms to the metastable bct10 structure through the bcc end-member. **B**, the scattering intensity profile modelled using the WPDFM method and lattice deformation matrix only (line) are compared with the average solution of the Debye scattering equation for a set of 100 equivalent crystals with random position of the shape relative to the lattice (dot). **C**, the scattering intensity profile modelled using the WPDFM method and both lattice and shape deformation tensors (line) are compared with the average solution of the Debye scattering equation for a set of 100 equivalent crystals with random position of the shape relative to the lattice (dot). The independent global optimization of the Debye profiles modelled with the WPDFM from fcc 8nm cubic crystal yielded the estimation of the true parameters with an error < 0.5% and GoF < 0.01.

### 3.6. Phase-Transition–Induced Shape Deformation in a Solid–Solid Transformation

The solid–solid phase–transition–driven shape evolution of Cu–Pd nanocrystals during thermal annealing provides an ideal test case for assessing the importance of explicitly including shape and lattice deformation in diffraction-profile modelling. In this system, the transformation from the disordered *fcc* A1 random alloy to the ordered *bcc* B2 intermetallic phase is accompanied by a measurable and systematic deformation of particle shape across the ensemble. This makes it possible to extract both the lattice transformation and the induced shape deformation directly from the diffraction signal. The crystallite-size distribution is first determined reliably in the A1 state; the subsequent deformation occurring during the A1→B2 transformation can then be refined directly from the PD profiles or imposed as a physically motivated constraint to stabilise the refinement of thermal and mechanical disorder parameters.

To isolate shape-evolution effects from other microstructural processes, such as coalescence, grain growth, or melting, classical MD simulations were employed as a controlled surrogate for real experimental samples. Spherical nanocrystals with a size ranging from 10 nm to 30 nm were equilibrated at room temperature. In the A1 phase, Cu and Pd atoms were distributed randomly, whereas in the B2 phase, atoms were arranged in an ideal intermetallic ordering. Debye-scattering–based powder profiles were then generated by summing the intensities computed from the final equilibrium MD snapshots, naturally incorporating thermal (dynamic) disorder and static or mechanical distortions such as surface relaxation. These MD-Debye profiles present modelling challenges similar to experimental datasets, yet without instrument broadening, noise, or background, artefacts that in practice can often be mitigated in high-resolution, high-quality sample measurements.

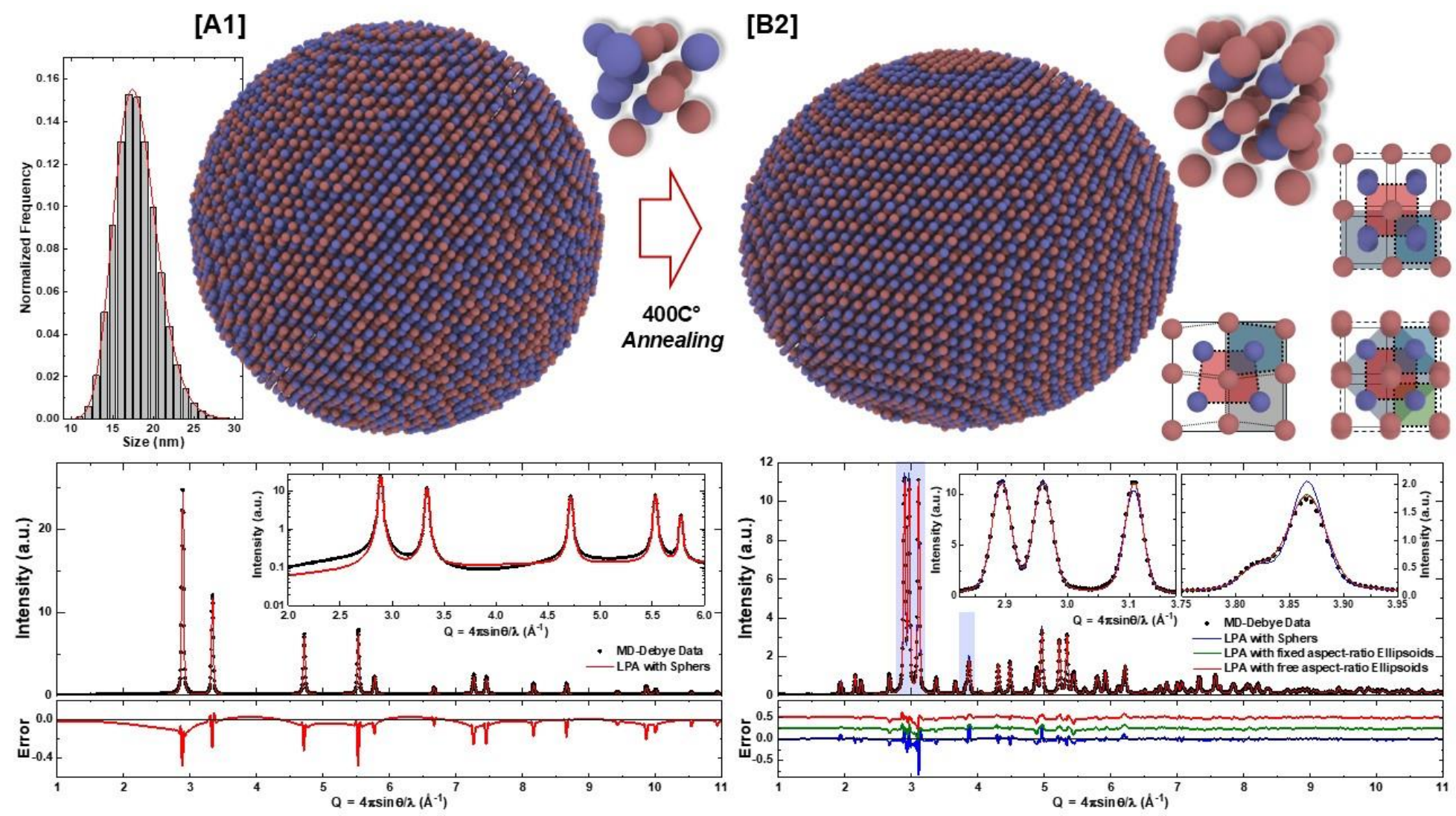

**Figure 9** Modelling of the Cu–Pd random-alloy (A1) to intermetallic (B2) transformation. Molecular dynamics (MD) equilibrium configurations were generated to reproduce the powder-scattering signatures of the two end members of the thermal-annealing pathway in polydisperse Cu–Pd nanocrystal powders (Wang *et al.*, 2016). Virtual-experiment diffraction profiles were computed as weighted sums of intensities obtained from the solution of the Debye Scattering Equation for spherical MD crystals with sizes ranging from 10 to 30 nm with lognormal frequency. Given that single MD snapshots were used, the models naturally include both thermal disorder and static/mechanical distortion (e.g., surface relaxation). The resulting synthetic profiles were analysed using the Whole Pair Distribution Function Modelling (WPDFM) framework, optimising via simulated annealing the following parameters: lattice constants; particle-size distribution; thermal disorder (isotropic Debye–Waller factor and short-range correlation length); effective mechanical disorder (Wilkens–Krivoglaz dislocation model); and shape-distortion modes describing stretch and shear of the B2 intermetallic phase. The optimized model's parameters are listed in **Table 4**. Notably, in the intermetallic phase, the atomic positions exhibit systematic displacements from the ideal body-centred cubic arrangement, necessitating a primitive orthorhombic description with a unit cell four times larger than the parent A1 *fcc*.

During MD equilibration, the *fcc* intermetallic phase progressively develops a B2-like ordering (see **Table 3**) accompanied by a change in particle shape from spherical to ellipsoidal.

Representative configurations and corresponding PD intensity profiles are shown in **Figure 9**. Although the final configurations display the expected B2-type ordering and are free of crystallographic defects (dislocations, stacking faults, twins, etc.), they also exhibit systematic periodic displacements from ideal *bcc* positions, producing a Bragg-peak sequence consistent with an array of eight B2-like building blocks forming a primitive orthorhombic supercell. This requires a four-fold enlargement of the conventional A1 *fcc* cell to fully describe the resulting symmetry reduction.

**Table 3** **Cu–Pd B2-like structure: fractional atomic positions in the orthorhombic lattice.** Independent WPDFM refinement of the MD–Debye profile returns fractional coordinates consistent with no deviations from the expected site values.

| Site ID | Element Type | x | y | z |
|---|---|---|---|---|
| 1 | Cu | 0.0000 | 0.0000 | 0.0000 |
| 2 | Cu | 0.5600 | 0.0000 | 0.0000 |
| 3 | Cu | 0.0000 | 0.5000 | 0.0000 |
| 4 | Cu | 0.5600 | 0.5000 | 0.0000 |
| 5 | Cu | 0.0600 | 0.0000 | 0.5000 |
| 6 | Cu | 0.5000 | 0.0000 | 0.5000 |
| 7 | Cu | 0.0600 | 0.5000 | 0.5000 |
| 8 | Cu | 0.5000 | 0.5000 | 0.5000 |
| 9 | Pd | 0.2800 | 0.2500 | 0.2260 |
| 10 | Pd | 0.7800 | 0.2500 | 0.2740 |
| 11 | Pd | 0.2800 | 0.7500 | 0.2260 |
| 12 | Pd | 0.7800 | 0.7500 | 0.2740 |
| 13 | Pd | 0.2800 | 0.2500 | 0.7740 |
| 14 | Pd | 0.7800 | 0.2500 | 0.7260 |
| 15 | Pd | 0.2800 | 0.7500 | 0.7740 |
| 16 | Pd | 0.7800 | 0.7500 | 0.7260 |

The MD-Debye PD profiles were analysed using the WPDFM framework, refining lattice parameters, size distributions, thermal disorder (modelled via an isotropic Debye–Waller factor ($B_{iso}$) with short-range correlation length $K_m$), mechanical disorder (Wilkens–Krivoglaz model, (Wilkens, 1970, 1969)), and B2-phase shape-deformation modes (elongation and shear). The size distributions refined from the A1 profiles agree closely with the nominal lognormal distribution used to generate the MD ensemble. Discrepancies arise because the approximation in the isotropic lattice distortion (BuILD) thermal model (Leonardi & Leoni, 2023). The diffuse scattering of the random alloy cannot be fully captured by a single-site type isotropic thermal model (i.e., $B_{iso}(Cu) = B_{iso}(Pd)$ and $K_m(Cu) = K_m(Pd)$): in the A1 state, Cu and Pd atoms occupy equivalent crystallographic sites but possess different phonon spectra, which the mixed-site isotropic thermal model cannot discriminate. Using an anisotropic thermal model improves the agreement but does not fully eliminate the mismatch, confirming that the thermal model, not the size-distribution extraction, is the principal limiting factor in the A1 case. In contrast, modelling the ordered B2-like phase is less affected by this limitation. Because Cu and Pd occupy distinct crystallographic sites, their differing vibrational behaviour, and thus their distinct phonon structures, are captured naturally by the thermal model and the independent $B_{iso}(Cu) \neq B_{iso}(Pd)$ and $K_m(Cu) \neq K_m(Pd)$ values. As a result, the agreement between the MD-Debye patterns and the WPDFM refinement is markedly improved for the B2-like structures. Crucially, the WPDFM implementation incorporates thermal diffuse scattering (TDS) implicitly through the short-range correlation length in the thermal disorder model. Consequently, no additional or dedicated TDS model is required when particle shape deforms: the TDS contribution automatically adjusts to the deformed shape through the combined directional-Fourier formalism.

The B2-like phase was refined using the size distribution extracted from the A1 state. Two strategies were compared: (i) assuming spherical particles, and (ii) allowing ellipsoidal particles with either constrained (*fcc→bcc*-like) or freely optimised aspect ratios and shear deformation (**Table 4**). Allowing for shape deformation improves the goodness-of-fit and $\chi^2$ by ~10% relative to the spherical assumption (Rietveld, 1969). Visual inspection shows that the elliptically deformed model reproduces peak intensities, especially peak maxima, far more accurately than the spherical model, while only marginally altering peak-broadening characteristics.

**Table 4 Optimized powder-diffraction model parameters for the Cu–Pd phase-transition end members.** Italicized values denote fixed model coefficients. The Goodness of Fit follows its standard Poisson-weighted definition, $GoF = \sqrt{[\sum (I_{model} - I_{obs})^2/I_{obs}]/[N-M]}$, where $N$ is the number of data points in the $Q$-range 1.0–11.0 Å⁻¹ and $M$ the number of refined degrees of freedom. The shape-deformation tensor was constrained to have unit determinant so that the deformation does not alter the crystallite volume within the modelled powder ensemble. A single thermal parameter is shared by Cu and Pd atoms in the case of A1 random alloy configuration.

| Model Parameter | | A1 random alloy | B2-like intermetallic | | |
|---|---|---|---|---|---|
| Shape | | Sphere | Sphere | Ellipsoid fixed aspect-ratio | Ellipsoid free aspect-ratio |
| $T_{s,11}$ | | *1.0* | *1.0* | *1.075517* | 1.104586 |
| $T_{s,22}$ | | *1.0* | *1.0* | *1.075517* | 1.070831 |
| $T_{s,33}$ | | *1.0* | *1.0* | *0.864501* | 0.854720 |
| $T_{s,12} = T_{s,21}$ | | *0.0* | *0.0* | *0.0* | -0.066871 |
| $T_{s,13} = T_{s,31}$ | | *0.0* | *0.0* | *0.0* | -0.050587 |
| $T_{s,23} = T_{s,32}$ | | *0.0* | *0.0* | *0.0* | 0.066413 |
| $B_{iso}$ (Å²) | Cu | 0.977976 | 0.493633 | 0.493430 | 0.486070 |
| | Pd | | 0.546028 | 0.543710 | 0.545179 |
| $K_m$ (Å) | Cu | 1.071598 | 1.867679 | 2.194552 | 2.281078 |
| | Pd | | 2.571774 | 2.419139 | 1.865144 |
| ρ | | 4.00 × 10⁻⁶ | 7.47 × 10⁻⁶ | 3.56 × 10⁻⁶ | 6.40 × 10⁻⁶ |
| Re (nm) | | 9.82 | 8.90 | 6.74 | 9.00 |
| mix | | 0.991617 | 0.998377 | 0.998283 | 0.991541 |
| $GoF$ | | 18.7455 | 11.7386 | 10.1448 | 10.0701 |
| $\chi^2$ | | 700327.8 | 274762.7 | 205216.4 | 201596.9 |

A1 optimized lattice parameters: a = b = c = 3.768742Å, α = β = γ = 90°
B2-like optimized lattice parameters: a = 5.83744Å, b = 5.60361Å, c = 6.50065Å, α = β = γ = 90°
Size distribution: lognormal mean = 18.00 nm, s.d. = 2.75 nm

The freely refined aspect ratio and shear parameters of the ellipsoids are consistent with those expected from the orthorhombic distortion of the B2-like supercell, demonstrating that even modest shape deformations can be reliably extracted from the diffraction data. Moreover, refining the aspect ratio improves the estimation of the thermal short-range correlation length, bringing it closer to the value obtained for the A1 state. Mechanical disorder parameters remain broadly consistent across models, though the ellipsoidal particles show a slightly higher effective dislocation density, which is consistent with the enhanced static/mechanical disorder expected when increased surface-to-volume ratios amplify surface-relaxation effects. Consistent with experimental observations reported in the literature (Scardi & Gelisio, 2016; Gelisio *et al.*, 2013; Scardi *et al.*, 2015), the lattice distortion arising from bond relaxation at the free surfaces of nanocrystals is best described, within the Wilkens–Krivoglaz formalism, by edge-type dislocations. Indeed, as reported in **Table 4**, a mixing coefficient close to unity signifies that the anisotropic strain field is almost entirely associated with edge dislocation character, rather than a mixed or screw-type contribution, which corresponds to a mixing coefficient of zero.

To assess the feasibility of extracting deformation parameters under realistic experimental conditions, synthetic Gaussian noise was added to the virtual Debye profiles, with a standard deviation proportional to √I, mimicking counting statistics. The noisy profiles were then refined with all deformation parameters allowed to vary freely. The results (**Table 5**) show only minimal deviations compared to the optimisation of the noise-free datasets (**Table 4**). Although the introduction of noise leads to an expected increase in the fit residuals, with larger noise levels yielding higher overall errors, the refined structural parameters remain largely stable, with noticeable variations occurring only at the highest noise amplitudes. Importantly, the shape deformation coefficients exhibit weak sensitivity to noise, confirming the robustness of their extraction under conditions representative of real experimental data. Other instrumental aberrations can be effectively mitigated through appropriate experimental design and data treatment. For instance, statistical noise can be reduced by employing sufficiently long exposure times, while instrumental broadening can be minimised using high-resolution powder diffraction beamlines at synchrotron sources, where it can often be considered negligible relative to the data collection step size. Similarly, background contributions arising from the sample environment or holder can be reliably removed through careful instrumental calibration procedures, as routinely performed in pair distribution function (PDF) experiments (Billinge & Farrow, 2013; Egami & Billinge, 2003).

**Table 5 Noise contribution to optimized powder-diffraction model parameters for the B2-like Cu–Pd end member .** The Goodness of Fit follows its standard Poisson-weighted definition, $GoF = \sqrt{[\sum (I_{model} - I_{obs})^2/I_{obs}]/[N-M]}$ , where $N$ is the number of data points in the $Q$-range 1.0–11.0 Å$^{-1}$ and $M$ the number of refined degrees of freedom. The shape-deformation tensor was constrained to have unit determinant so that the deformation does not alter the crystallite volume within the modelled powder ensemble. The noise was sampled from a normal distribution with a standard deviation $\xi\sqrt{I_{obs}}$.

| Model Parameter | Noise $\xi = 1.0$ | Noise $\xi = 2.0$ | Noise $\xi = 5.0$ |
|---|---|---|---|
| $T_{s,11}$ | 1.104586 | 1.104586 | 1.099906 |
| $T_{s,22}$ | 1.070831 | 1.070831 | 1.077248 |

| | | | | |
|---|---|---|---|---|
| $T_{s,33}$ | | 0.854720 | 0.854720 | 0.860499 |
| $T_{s,12} = T_{s,21}$ | | -0.066871 | -0.066871 | -0.128027 |
| $T_{s,13} = T_{s,31}$ | | -0.050587 | -0.050587 | -0.053559 |
| $T_{s,23} = T_{s,32}$ | | 0.066413 | 0.066413 | 0.053235 |
| $B_{iso}$ (Å$^2$) | Cu | 0.486070 | 0.486070 | 0.478979 |
| | Pd | 0.545179 | 0.545179 | 0.547909 |
| $K_m$ (Å) | Cu | 2.281078 | 2.281078 | 2.335296 |
| | Pd | 1.865144 | 1.865144 | 2.130952 |
| ρ | | $6.40 \times 10^{-6}$ | $6.40 \times 10^{-6}$ | $6.56 \times 10^{-6}$ |
| Re (nm) | | 9.00 | 9.00 | 8.35 |
| mix | | 0.991541 | 0.991541 | 0.987452 |
| $GoF$ | | 10.1064 | 10.1873 | 10.6468 |
| $\chi^2$ | | 203053.0 | 206316.8 | 225348.4 |

B2-like optimized lattice parameters: a = 5.83744Å, b = 5.60361Å, c = 6.50065Å, α = β = γ = 90°. Size distribution: lognormal mean = 18.00 nm, s.d. = 2.75 nm

A common outcome of all refinements is that the effective dislocation radius $R_{\mathrm{e}}$is comparable to or smaller than the minimum crystallite size present in the size distribution. This behaviour can be rationalized by noting that the same distortion model is applied uniformly across the entire ensemble of particles. When $R_{\mathrm{e}}$ is smaller than the crystallite dimensions, the edge-dislocation model retains physical meaning and ensures a finite elastic energy integrated over the crystal volume. In contrast, the values of $R_{\mathrm{e}}$ exceeding the smallest particle size would imply that the strain field does not relax as the dislocation approaches the surface of the smaller crystallites, leading to an unphysical description of the distortion. Notably, when the transformation matrix is refined without constraints, the rotational components are non-zero (see **Appendix B** for details). In particular, the $T_{23}$ term, although relatively small in magnitude, is responsible for a tilt of the $c$ axis of approximately 15deg with respect to the Cartesian reference frame of the parent structure base unit cell. In contrast, the combined effect of all off-diagonal components produces additional rotations of the $a$ and $b$ axes within the basal plane that do not exceed about 2deg.

While the in-plane rotations are marginal, the pronounced tilt of the $c$ axis clearly illustrates the advantage of the methodology we introduced. Indeed, detecting such a shape-deformation tilt would be extremely challenging using transmission electron microscopy alone, as well as other currently available PD line profile analysis methods, underscoring the sensitivity of the diffraction-based approach. Overall, this realistic example demonstrates that explicitly accounting for shape deformation is essential to correctly interpret diffraction signals across solid–solid phase transformations. By enabling coupled refinement of lattice distortion, shape evolution, thermal disorder, and mechanical strain, the proposed framework provides a robust route for analysing real nanostructured materials undergoing structural transitions.

### 4. Conclusions

In this work, we examined the effects of general linear transformations, including deformation and rotation, applied independently or jointly to both the lattice and the external shape of crystalline domains. Building on these concepts, we introduced a unified formalism based on the construction of a virtual lattice, which allows known analytical Fourier coefficients of undeformed reference shapes to be retained while consistently mapping arbitrary shape and lattice transformations into reciprocal space. This strategy enables the treatment of general domain geometries with arbitrary orientation relative to the crystal lattice, without requiring the derivation of new analytical shape functions for each specific deformation scenario.

The proposed formalism greatly expands the modelling capabilities of existing, well-established tools for powder-diffraction line-profile analysis, such as TOPAS, by lifting the traditional constraints imposed by *a priori* predefined CVFs. By enabling the dynamic deformation of the crystalline domain shape and/or the underlying lattice, the approach allows both effects to be treated as active and refinable degrees of freedom. Most notably, it enables the refinement of the relative orientation between shape and structure bases, a capability that was previously inaccessible within conventional line-profile modelling frameworks.

We demonstrated that this formalism integrates naturally within established powder-diffraction modelling strategies in both reciprocal and real space, including advanced line-profile and total-scattering approaches. The resulting diffraction and PDF profiles faithfully reproduce the combined effects of size anisotropy, lattice distortion, and coupled shape–lattice transformations across a wide range of nanocrystal morphologies. A key aspect of the virtual-lattice construction is that a single deformation tensor governs both Bragg and diffuse scattering contributions, including thermal diffuse scattering, ensuring internal consistency without the introduction of additional *ad hoc* modelling terms.

Through extensive validation against virtual diffraction and total-scattering datasets generated using the Debye scattering equation, we showed that the proposed methodology allows deformation parameters to be retrieved reliably from powder diffraction data alone. Even in cases where simplified thermal or mechanical distortion models limit a full description of the total-scattering signal, such as in random alloys, where species-dependent phonon spectra cannot be resolved with known models, the extraction of shape and lattice deformation remains robust. For ordered intermetallic structures, where atomic species occupy distinct crystallographic sites, the framework captures the coupled evolution of shape and lattice with high fidelity. In this context, we show that while in-plane rotations of the particle axes are relatively small, a pronounced tilt of the *c* axis can be resolved unambiguously. Detecting such a shape-induced tilt would be extremely challenging using transmission electron microscopy alone, as well as with currently available powder-diffraction line-profile methods that rely on predefined shape models and fixed shape–lattice orientation assumptions. This result highlights the sensitivity of the diffraction-based approach introduced here and its ability to resolve deformation features that were previously inaccessible to ensemble-averaged techniques.

Overall, the formalism introduced here provides a general and flexible route for modelling diffraction profiles from realistic nanostructured materials, explicitly treating domain shape and lattice as independent yet coupled degrees of freedom. By enabling deformation-aware analysis without imposing crystallographic constraints on the external morphology, this framework extends the interpretive power of powder-diffraction and total-scattering methods. It opens the door to more accurate size-, shape-, and deformation-resolved studies of nanocrystals, with direct relevance for investigating phase transformations, strain-driven distortions, anisotropic growth, and morphology evolution in functional nanomaterials.

**Acknowledgements** Saudi Aramco is acknowledged for the continuous support and for the permission to publish this manuscript. This research was supported in part by Lilly Endowment, Inc., through its support for the Indiana University Pervasive Technology Institute.

**Conflicts of interest** The authors declare no conflicts of interest. The funders had no role in the design of the study; in the collection, analyses, or interpretation of data; in the writing of the manuscript; or in the decision to publish the results.

**Data availability** The original contributions presented in this study are included in the article/supplementary material. Further inquiries can be directed to the corresponding authors.

### References

Allegra, G. & Wilson, A. J. C. (1983). *Acta Crystallographica A* **39**, 280–282.

Aviles, K. M. & Lear, B. J. (2025). *ACS Nanoscience Au* **5**, 117–127.

Le Bail, A., Duroy, H. & Fourquet, J. L. (1988). *Mater. Res. Bull.* **23**, 447–452.

Balic-Zunic, T. & Dohrup, J. (1999). *Powder Diffr.* **14**, 203–207.

Billinge, S. J. L. & Farrow, C. L. (2013). *Journal of Physics: Condensed Matter* **25**, 454202.

Boukouvala, C., Daniel, J. & Ringe, E. (2021). *Nano Converg.* **8**, 26.

Brandstetter, S., Derlet, P. M., Van Petegem, S. & Van Swygenhoven, H. (2008). *Acta Mater.* **56**, 165–176.

Cho, M. G., Sytwu, K., Rangel DaCosta, L., Groschner, C., Oh, M. H. & Scott, M. C. (2024). *ACS Nano* **18**, 29736–29747.

Coelho, A. A. (2018). *J. Appl. Crystallogr.* **51**, 210–218.

Crisp, R. W. & Singh, S. (2024). *Cryst. Growth Des.* **24**, 7739–7741.

Debye, P. (1913). *Ann. Phys.* **348**, 49–92.

Debye, P. (1915). *Nachrichten von Der Koeniglichen Gesellschaft Der Wissenschaften Zu Goettingen, Mathematisch-Physikalische Klasse* **1915**, 70–76.

Du, C. X., van Anders, G., Newman, R. S. & Glotzer, S. C. (2017). *Proceedings of the National Academy of Sciences* **114**, E3892–E3899.

Ectors, D., Goetz-Neunhoeffer, F. & Neubauer, J. (2015). *J. Appl. Crystallogr.* **48**, 189–194.

Edgar C. Bain (1924). *Transactions of the American Institute of Mining and Metallurgical Engineers* **70**, 25–46.

Egami, T. & Billinge, S. J. L. (2003). Undeneath the Bragg peaks. Structural Analysis of Complex Materials.

Farrow, C. L., Juhas, P., Liu, J. W., Bryndin, D., Božin, E. S., Bloch, J., Proffen, T. & Billinge, S. J. L. (2007). *Journal of Physics: Condensed Matter* **19**, 335219.

Fichthorn, K. A. & Yan, T. (2021). *The Journal of Physical Chemistry C* **125**, 3668–3679.

Gelisio, L., Beyerlein, K. R. & Scardi, P. (2013). *Thin Solid Films* **530**, 35–39.

Giacovazzo, Carmelo. (2011). Fundamentals of crystallography Oxford University Press.

Ino, T. & Minami, N. (1979). *Acta Crystallographica A* **35**, 163–170.

Khorsand Zak, A. & Hashim, A. M. (2025). *Sci. Rep.* **16**, 1717.

Krawitz, A. D. . (2001). Introduction to diffraction in materials, science, and engineering John Wiley.

Leonardi, A. (2021). *IUCrJ* **8**, 257–269.

Leonardi, A., Beyerlein, K. R., Xu, T., Li, M., Leoni, M. & Scardi, P. (2011). *Zeitschrift Für Kristallographie Proceedings* **1**, 37–42.

Leonardi, A. & Bish, D. L. (2016). *J. Appl. Crystallogr.* **49**, 1593–1608.

Leonardi, A. & Leoni, M. (2023). *Acta Mater.* **243**, 118491.

Leonardi, A., Leoni, M. & Scardi, P. (2013). *J. Appl. Crystallogr.* **46**, 63–75.

Leonardi, A., Leoni, M., Siboni, S. & Scardi, P. (2012). *J. Appl. Crystallogr.* **45**, 1162–1172.

Leonardi, A., Neder, R. & Engel, M. (2022). *J. Appl. Crystallogr.* **55**, 329–339.

Leoni, M. (2014). *TOPAS 5 Technical Reference, User Manual* 73.

Leoni, M. (2019*a*). *International Tables for Crystallography Vol. H* 524–537.

Leoni, M. (2019*b*). Vol. *H*, *International Tables for Crystallography*. pp. 524–537.

Leoni, M., Confente, T. & Scardi, P. (2006). *Zeitschrift Fur Kristallographie, Supplement* **1**, 249–254.

Li, C., Clament Sagaya Selvam, N. & Fang, J. (2023). *Nanomicro Lett.* **15**, 83.

Loopstra, B. O. & Rietveld, H. M. (1969). *Acta Crystallogr. B* **25**, 787–791.

Marks, L. D. & Peng, L. (2016). *Journal of Physics: Condensed Matter* **28**, 053001.

Minami, N. & Ino, T. (1979). *Acta Crystallographica Section A* **35**, 171–176.

Mohan, A. C., Athira, A., Nair, B. P., Sivasubramanian, G., Sreekanth, K. M., Anoop, G., Sree, S. P. & Sreedhar, K. M. (2024). *Sci. Rep.* **14**, 32067.

National Science Foundation, 2024 https://www.nsf.gov/mps/updates/tiny-crystals-huge-impact-nsf-launches-new-nanocrystals.

Neumann, S. & Rafaja, D. (2024). https://doi.org/10.3390/powders.

Nguyen, Q. N., Chen, R. & Xia, Y. (2023). *Trends Chem.* **5**, 748–762.

Pawley, G. S. (1981). *J. Appl. Crystallogr.* **14**, 357–361.

Popa, N. C. (1998). *J. Appl. Crystallogr.* **31**, 176–180.

Rietveld, H. M. (1967). *Acta Crystallogr.* **22**, 151–152.

Rietveld, H. M. (1969). *J. Appl. Crystallogr.* **2**, 65–71.

Scardi, P. (2020). *Cryst. Growth Des.* **20**, 6903–6916.

Scardi, P., Azanza Ricardo, C. L., Perez-Demydenko, C. & Coelho, A. A. (2018). *J. Appl. Crystallogr.* **51**, 1752–1765.

Scardi, P. & Gelisio, L. (2016). *Sci. Rep.* **6**, 22221.

Scardi, P., Leonardi, A., Gelisio, L., Suchomel, M. R., Sneed, B. T., Sheehan, M. K. & Tsung, C.-K. (2015). *Phys. Rev. B* **91**, 155414.

Scardi, P. & Leoni, M. (2001). *Acta Crystallographica A* **57**, 604–613.

Scardi, P. & Leoni, M. (2002). *Acta Crystallographica A* **58**, 190–200.

Scardi, P., Leoni, M. & Beyerlein, K. R. (2011). *Zeitschrift Für Kristallographie* **226**, 924–933.

Scardi, P., Leoni, M. & Delhez, R. (2004). *J. Appl. Crystallogr.* **37**, 381–390.

Setyawan, W. & Curtarolo, S. (2010). *Comput. Mater. Sci.* **49**, 299–312.

Stephens, P. W. (1999). *J. Appl. Crystallogr.* **32**, 281–289.

Stokes, A. R. & Wilson, A. J. C. (1942). *Mathematical Proceedings of the Cambridge Philosophical Society* **38**, 313–322.

Ungár, T., Révész, Á. ;. & Borbély, A. (1998). *J. Appl. Crystallogr.* **31**, 554–558.

Urban, J. P. (1975). *Acta Crystallogr. A* **31**, 95–100.

Various (2005). *International Tables for Crystallography*.

Waitz, T., Tsuchiya, K., Antretter, T. & Fischer, F. D. (2009). *MRS Bull.* **34**, 814–821.

Wang, C., Chen, D. P., Sang, X., Unocic, R. R. & Skrabalak, S. E. (2016). *ACS Nano* **10**, 6345–6353.

Warren, B. E. (1990). X-Ray Diffraction Mineola, N. Y.: Dover.

Warren, B. E. & Averbach, B. L. (1950). *J. Appl. Phys.* **21**, 595–599.

Warren, B. E. & Averbach, B. L. (1952). *J. Appl. Phys.* **23**, 497–497.

Wilkens, M. (1969). Vol. *Fundamental Aspects of Dislocation Theory*, edited by J. A. Simmons & R. Bullough. pp. 1195–1221. Washington, D.C.: National Bureau of Standards.

Wilkens, M. (1970). *Physica Status Solidi* **2**, 359–370.

Wilson, A. J. C. (1955). *Il Nuovo Cimento* **1**, 277–283.

Wilson, A. J. C. (1962). X-Ray Optics.

Xia, Y., Xiong, Y., Lim, B. & Skrabalak, S. E. (2009). *Angew. Chem. Int. Ed.* **48**, 60–103.

Yakovlev, E. V., Chaudhuri, M., Kryuchkov, N. P., Ovcharov, P. V., Sapelkin, A. V. & Yurchenko, S. O. (2019). *J. Chem. Phys.* **151**, 114502.

Yurchenko, S. O. (2014). *J. Chem. Phys.* **140**, 134502.

Yurchenko, S. O., Kryuchkov, N. P. & Ivlev, A. V. (2015). *J. Chem. Phys.* **143**, 034506.

Yurchenko, S. O., Kryuchkov, N. P. & Ivlev, A. V. (2016). *Journal of Physics: Condensed Matter* **28**, 235401.

### Appendix A. Tensor matrix

A given Bravais lattice $\Lambda(\boldsymbol{r})$ with basis $\boldsymbol{A} = (\boldsymbol{a} \quad \boldsymbol{b} \quad \boldsymbol{c})$ is described by the metric tensor:

$$\mathbf{G} = \begin{pmatrix} \boldsymbol{a}\cdot\boldsymbol{a} & \boldsymbol{a}\cdot\boldsymbol{b} & \boldsymbol{a}\cdot\boldsymbol{c} \\ \boldsymbol{b}\cdot\boldsymbol{a} & \boldsymbol{b}\cdot\boldsymbol{b} & \boldsymbol{b}\cdot\boldsymbol{c} \\ \boldsymbol{c}\cdot\boldsymbol{a} & \boldsymbol{c}\cdot\boldsymbol{b} & \boldsymbol{c}\cdot\boldsymbol{c} \end{pmatrix} = \begin{pmatrix} a^2 & ab\cos\gamma & ac\cos\beta \\ ab\cos\gamma & b^2 & bc\cos\alpha \\ ac\cos\beta & bc\cos\alpha & c^2 \end{pmatrix}, \quad (A.1)$$

such that,

$$\det(\mathbf{G}) = V^2 = a^2b^2c^2(\sin^2\alpha - \cos^2\beta - \cos^2\gamma + 2\cos\alpha\cos\beta\cos\gamma), \quad (A.2)$$

where V is the volume of the unit cell. Then, a column vector $\boldsymbol{n} = (h \quad k \quad l)^T$ describes in the reciprocal space the normal direction to a given crystallographic plane (*hkl*). The corresponding interplanar spacing is:

$$d_{\boldsymbol{n}} = (d_{\boldsymbol{n}}^*)^{-1} = (\boldsymbol{n}^T \cdot \mathbf{G}^{-1} \cdot \boldsymbol{n})^{-1/2} = V \begin{pmatrix} (hbc\sin\alpha)^2 + (kac\sin\beta)^2 + (lab\sin\gamma)^2 + \\ -2hkabc^2(\cos\gamma - \cos\alpha\cos\beta) + \\ -2hlab^2c(\cos\beta - \cos\alpha\cos\gamma) + \\ -2kla^2bc(\cos\alpha - \cos\beta\cos\gamma) \end{pmatrix}^{-1/2}. \quad (A.3)$$

The Cholesky decomposition of the metric tensor $\mathbf{G} = \mathbf{M}\cdot\mathbf{M}^{\mathrm{T}}$ provides the matrix $\mathbf{M}$:

$$\mathbf{M} = \begin{pmatrix} a & 0 & 0 \\ b\cos\gamma & b\sin\gamma & 0 \\ c\cos\beta & \frac{c}{\sin\gamma}(\cos\alpha - \cos\beta\cos\gamma) & \frac{V_{cell}}{ab\sin\gamma} \end{pmatrix}, \quad (A.4)$$

$$\det(\mathbf{M}) = V = abc\sqrt{\sin^2\beta\sin^2\gamma - (\cos\alpha - \cos\beta\cos\gamma)^2}, \quad (A.5)$$

that allows the representation in orthogonal coordinates $\boldsymbol{n}_\perp$ of a given vector n in reciprocal space:

$$\boldsymbol{n}_\perp = \mathbf{M}^{-1}\cdot\boldsymbol{n}. \quad (A.6)$$

### Appendix B. Deformation tensors

Deformations of a body are most generally described using second-rank tensors. A triaxial deformation, incorporating simultaneous changes in length, shape, and orientation, is represented by the deformation tensor, which in its general form can be written as:

$$\mathbf{T} = \begin{pmatrix} T_{xx} & T_{xy} & T_{xz} \\ T_{yx} & T_{yy} & T_{yz} \\ T_{zx} & T_{zy} & T_{zz} \end{pmatrix}, \quad (A.7)$$

We can diagonalize **T** to find the principal axes, i.e. the directions in which the deformation does not cause any shear. This corresponds to finding the eigen-decomposition:

$$\mathbf{T} = \mathbf{R}^{-1} \cdot \begin{pmatrix} T_x & 0 & 0 \\ 0 & T_y & 0 \\ 0 & 0 & T_z \end{pmatrix} \cdot \mathbf{R} = \mathbf{R}^{-1} \cdot \mathbf{T}_t \cdot \mathbf{R}, \quad (A.8)$$

From a computational standpoint, $T_x$, $T_y$, and $T_z$correspond to the eigenvalues of **T**, while the columns of **R** are the associated eigenvectors.

The matrix **R** represents the rotation that must be applied to the fixed reference frame to align it with the principal axes of the deformation. This rotation can equivalently be parametrized using three Euler angles $(\chi, \psi, \varphi)$, expressed as the product of the corresponding elementary Euler rotation matrices about three independent axes. Introducing the matrices that describe a rotation by a generic angle $\vartheta$ about the **X** and **Z** axes of the fixed frame, we write:

$$\mathbf{X}_\vartheta = \begin{pmatrix} 1 & 0 & 0 \\ 0 & cos\,\vartheta & -sin\,\vartheta \\ 0 & sin\,\vartheta & cos\,\vartheta \end{pmatrix}, \text{ and } \mathbf{Z}_\vartheta = \begin{pmatrix} cos\,\vartheta & -sin\,\vartheta & 0 \\ sin\,\vartheta & cos\,\vartheta & 0 \\ 0 & 0 & 1 \end{pmatrix}. \quad (A.9)$$

Considering the customary Euler sequence consisting of three successive rotations about the fixed frame **Z**, **X**, and **Z** axes by the angles $\chi$, $\psi$, and $\varphi$, respectively, the resulting rotation matrix is obtained as:

$$\mathbf{R} = \mathbf{R}_{\chi,\psi,\varphi} = \mathbf{Z}_\chi \mathbf{X}_\psi \mathbf{Z}_\varphi =$$
$$\begin{pmatrix} \cos\chi\cos\varphi - \sin\chi\cos\psi\sin\varphi & -\sin\chi\cos\psi\cos\varphi - \cos\chi\sin\varphi & \sin\chi\sin\psi \\ \sin\chi\cos\varphi + \cos\chi\cos\psi\sin\varphi & -\sin\chi\sin\varphi + \cos\chi\cos\psi\cos\varphi & -\cos\chi\sin\psi \\ \sin\psi\sin\varphi & \sin\psi\cos\varphi & \cos\psi \end{pmatrix}, \quad (A.10)$$

where $\mathbf{R}^{-1}_{\chi,\psi,\varphi} = \mathbf{R}^{T}_{\chi,\psi,\varphi}$. Rotations are defined as active extrinsic Euler rotations in a right-handed orthonormal reference frame, applied successively about the fixed $Z_\phi$, $X_\psi$, and $Z_\chi$ axes. Using this rotation matrix, the full deformation tensor can be constructed bottom-up from the specified deformation parameters and rotation angles. This provides a convenient way to generate a deformed geometry, such as a distorted lattice or reshaped domain, and subsequently orient it relative to the initial reference frame.

### Appendix C. Powder diffraction profile modelling

Excluding defects and strain, each domain in a powder of a monoatomic material can be regarded as a finite portion of an infinite Bravais lattice. The underlying ideal lattice can be expressed as the Dirac comb:

$$(\boldsymbol{y}) = \sum_{p_1=-\infty}^{\infty} \sum_{p_2=-\infty}^{\infty} \sum_{p_3=-\infty}^{\infty} \delta(\boldsymbol{y} - p_1\boldsymbol{a} - p_2\boldsymbol{b} - p_3\boldsymbol{c}) = \sum_{p_1=-\infty}^{\infty} \sum_{p_2=-\infty}^{\infty} \sum_{p_3=-\infty}^{\infty} \delta(\boldsymbol{y} - \boldsymbol{p}) \quad , \tag{A.11}$$

where $\mathbf{p} = p_1\mathbf{a} + p_2\mathbf{b} + p_3\mathbf{c}$. A finite crystalline domain of shape $\omega$ is obtained by selecting the lattice points lying within the region $\Omega$. This is achieved clipping the infinite lattice with the characteristic indicator function:

$$\sigma_{\Omega,\omega}(\boldsymbol{y}) = \begin{cases} 1, \boldsymbol{y} \in \Omega \\ 0, \boldsymbol{y} \notin \Omega \end{cases}. \tag{A.12}$$

To construct the real-space density of the finite crystallite, the clipped lattice is populated with a single-atom density $\rho(\mathbf{y})$, typically the electron density for X-ray scattering or nuclear cross-section for neutrons. The density $\rho_{\Omega,\omega}(\boldsymbol{y})$ of the domain $\Omega$ with shape $\omega$ is then generated by populating the clipped lattice via convolution:

$$\rho_{\Omega,\omega}(\boldsymbol{y}) = \rho(\boldsymbol{y}) \otimes \left[\Lambda(\boldsymbol{y}) \cdot \sigma_{\Omega,\omega}(\boldsymbol{y})\right]. \tag{A.13}$$

This expression ensures that the resulting density is atomistically discrete yet preserves the smooth shape termination imposed by $\sigma_{\Omega,\omega}$, which in turn guarantees charge neutrality and continuity at the crystal surface (Ino & Minami, 1979). These properties explain the high accuracy of Fourier-based line-profile analysis compared with more traditional approaches that impose shapes directly in reciprocal space.

The amplitude, $Y_{\Omega,\omega}(\boldsymbol{q})$, and intensity, $I_{\Omega,\omega}(\boldsymbol{q})$, are obtained from the Fourier transform of the domain density,

$$\begin{aligned} Y_{\Omega,\omega}(\boldsymbol{q}) &= \Im\left[\rho_{\Omega,\omega}(\boldsymbol{y})\right] \\ I_{\Omega,\omega}(\boldsymbol{q}) &= Y_{\Omega,\omega}(\boldsymbol{q}) Y_{\Omega,\omega}^{*}(\boldsymbol{q}) = \left|\Im\left[\rho_{\Omega,\omega}(\boldsymbol{y})\right]\right|^{2}, \end{aligned} \tag{A.14}$$

where $\Im[\cdot]$ is the Fourier operator and $\mathbf{q}$ the reciprocal-space coordinate, i.e., the scattering momentum. Applying the convolution theorem yields

$$I_{\Omega,\omega}(\boldsymbol{q}) = \left|\Im[\rho(\boldsymbol{y})]\left(\Im[\Lambda(\boldsymbol{y})] \otimes \Im\left[\sigma_{\Omega,\omega}(\boldsymbol{y})\right]\right)\right|^{2} = f^{2}(\boldsymbol{q})|P(\boldsymbol{q}) \otimes F_{\omega}(\boldsymbol{q})|^{2},$$

(A.15)

in which $f(\boldsymbol{q})$ is the atomic scattering factor (or nuclear scattering length for neutrons, $f(\boldsymbol{q}) = b$),

$$P(\boldsymbol{q}) = \sum_{h=-\infty}^{\infty} \sum_{k=-\infty}^{\infty} \sum_{l=-\infty}^{\infty} \delta(\boldsymbol{q} - \boldsymbol{v}), \tag{A.16}$$

is the reciprocal-lattice Dirac comb with basis $\boldsymbol{A}^* = (\boldsymbol{a}^* \quad \boldsymbol{b}^* \quad \boldsymbol{c}^*)$ and lattice node vectors $\boldsymbol{v} = h\boldsymbol{a}^* + k\boldsymbol{b}^* + l\boldsymbol{c}^*$ , and $F_\omega(\boldsymbol{q}) = \Im[\sigma_{\Omega,\omega}(\boldsymbol{y})]$ is the Fourier transform of the domain's shape characteristic indicator function. Then equation (A.15) becomes:

$$I_{\Omega,\omega}(\boldsymbol{q}) = f^2(\boldsymbol{q}) \sum_{h=-\infty}^{\infty} \sum_{k=-\infty}^{\infty} \sum_{l=-\infty}^{\infty} |F_\omega(\boldsymbol{q} - \boldsymbol{v})|^2. \tag{A.17}$$

In other words, $F_\omega(\mathbf{q})$ is the shape transform, which determines the peak profile around each reciprocal-lattice point.

Because only the neighbourhood of each reciprocal-lattice point contributes to the powder pattern, it is convenient to evaluate the shape transform along the radial direction $\hat{\boldsymbol{v}} = \boldsymbol{v}/|\boldsymbol{v}|$ and write $F_\omega(\boldsymbol{v} + L\hat{\boldsymbol{v}}) = F_\omega(\boldsymbol{v} - L\hat{\boldsymbol{v}}) = F_\omega(\hat{\boldsymbol{v}}, L)$ which corresponds geometrically to counting scattering pairs at separation $L$ along $\hat{\boldsymbol{v}}$ (Leonardi, 2021; Leonardi *et al.*, 2013). In the Stokes–Wilson "ghost'' construction, $F_\omega(\hat{\boldsymbol{v}}, L)$ equals the common volume between the object and a copy shifted by $L \cdot \hat{\boldsymbol{v}}$ (Leonardi *et al.*, 2012; Scardi & Leoni, 2001; Wilson, 1962; Leoni, 2019*a*; Stokes & Wilson, 1942). For mono-parametric shapes of size $D_\omega$ this yields a real function expressible as:

$$F_\omega(\hat{\boldsymbol{v}}, L) = A_\omega(\hat{\boldsymbol{v}}, L, D_\omega) = \sum_{j=0}^{3} H_{\omega,j}(\hat{\boldsymbol{v}}) \left(\frac{L}{D_\omega}\right)^j, 0 \le L < K_\omega(\hat{\boldsymbol{v}}) D_\omega, \tag{A.18}$$

and zero beyond the directional cutoff $K_\omega(\hat{\mathbf{v}})\, D_\omega$. The coefficients $H_{\omega,j}$ and the cutoff $K_\omega$ depend on both the crystallite shape $\omega$ and the direction of observation $\hat{\boldsymbol{v}}$ (Scardi & Leoni, 2001).

The powder diffraction pattern is finally constructed by averaging the three-dimensional intensity over the surface of the sphere $S_r$ where $|\, \mathbf{q} \,| = Q = r$ as:

$$I_\Omega(Q) = \frac{1}{4\pi r^2} \int_{|\mathbf{q}|=r} I_{\Omega,\omega}(\boldsymbol{q}) \partial S_r. \tag{A.19}$$

For a perfect lattice, this spherical average reduces to a sum over symmetry-equivalent directions, providing an efficient route to the one-dimensional powder profile. Physically, the reciprocal lattice dictates where scattering occurs, while the atomic density sets the intensities, and the crystallite shape determines the peak profile via $F_\omega$.

In practical computations, the averaging is often carried out using the tangent-plane approximation, which treats the integration sphere locally as flat. This approximation accelerates calculation and is accurate for narrow features in reciprocal space (large crystalline domains), but becomes progressively less valid for broader features (small crystalline domains). As a rough example, for crystallites of 10 nm diameter, the tangent-plane approximation introduces an error in peak width of less than 2% for Q > 2 Å$^{-1}$, but can reach 8% in the small-angle region (Q < 1 Å$^{-1}$). As expected, some differences appear at small diffraction angles where the radius of the integration sphere is small (Scardi *et al.*, 2011), especially for complex-shaped domains. These considerations motivate the directional formulation above, where the primary contribution arises along $\hat{\boldsymbol{v}}$, and the accuracy of the diffraction modelling depends on an appropriate representation of the shape transform in that direction.